\documentstyle[aps,prd,preprint,floats,epsf]{revtex}

\newcommand{\DD}    {\displaystyle}

\begin{document}
\draft

\tighten 
\preprint{\tighten\vbox{
                        \hbox{\hfil CLNS 99/1601}
                        \hbox{\hfil CLEO 99--01}
}}


\title{Hadronic Structure 
       in the Decay \boldmath $\tau^-\rightarrow\nu_\tau\pi^-\pi^0\pi^0$ \\
       and the Sign of the Tau Neutrino Helicity }


\author{The CLEO Collaboration}
 
\date{15 February 1999}

\maketitle


\begin{abstract} 
Based on a sample corresponding to $4.3\times 10^6$ produced  
$\tau$-pair events, we have studied hadronic dynamics in the 
decay $\tau^-\to \nu_\tau\pi^-\pi^0\pi^0$ in data recorded by 
the CLEO~II detector operating at the CESR $e^+e^-$ collider.  
The decay is dominated by the process 
$\tau^-\to \nu_\tau a_1^-(1260)$, with the $a_1^-$ meson decaying 
to three pions 
predominantly via the lowest dimensional (mainly $S$-wave) 
$a_1\to\rho^-\pi^0$ Born amplitude.  
From model-dependent fits to the Dalitz plot and angular 
observables in bins of $3\pi$ mass, 
we find significant additional contributions from amplitudes 
for $a_1$ decay to $\sigma\pi$, $f_0(1370)\pi$ and $f_2(1270)\pi$, 
as well as higher dimensional $a_1 \to \rho\pi$ and $\rho^\prime\pi$ 
amplitudes. 
Notably, the squared $\sigma \pi$ amplitude 
accounts for approximately 15$\%$ 
of the total $\tau^-\to\nu_\tau \pi^-\pi^0\pi^0$ 
rate in the models considered.  
The data are well described using couplings to 
these amplitudes that are independent of the $3\pi$ mass.  
These amplitudes also provide a good description for the 
$\tau^- \to \nu_\tau\pi^-\pi^+\pi^-$ Dalitz plot distributions.  
We have searched for additional contributions from 
$\tau^-\to\nu_\tau \pi^{\prime -}(1300)$.  
We place $90\%$ confidence level upper limits on 
the branching fraction for this channel  
of between $1.0\times 10^{-4}$ and $1.9\times 10^{-4}$, 
depending on the $\pi^\prime$ decay mode considered.
The $\pi^-\pi^0\pi^0$ mass spectrum is parametrized by a 
Breit-Wigner form 
with a mass-dependent width 
which is specified according to the results of the Dalitz plot fits 
plus an unknown coupling to an $a_1 \to K^*K$ amplitude.
From a $\chi^2$ fit using this parametrization, 
we extract the pole mass and width
of the $a_1$, as well as the magnitude of the $K^*K$ coupling.  
We have also investigated the impact of a possible contribution 
from the $a_1^\prime(1700)$ meson on this spectrum. 
Finally, exploiting the parity-violating angular asymmetry in 
$a_1\to 3\pi$ decay, we determine the signed value of the 
$\tau$ neutrino helicity to be 
$h_{\nu_\tau} = -1.02 \pm 0.13\,\mbox{(stat.)}\pm 0.03\,
\mbox{(syst.$+$model)}$, confirming the left-handedness of 
the $\tau$ neutrino.  
\end{abstract}


\pacs{PACS numbers: 13.25.Jx, 13.35.Dx, 14.40.Cs, 14.60.Fg, 14.60.Lm
      \vspace{0pt}\\
\begin{center}
Submitted to {\sl Physical Review D\vspace{0pt}}
\end{center}
}



\begin{center}
D.~M.~Asner,$^{1}$ A.~Eppich,$^{1}$ J.~Gronberg,$^{1}$
T.~S.~Hill,$^{1}$ D.~J.~Lange,$^{1}$ R.~J.~Morrison,$^{1}$
H.~N.~Nelson,$^{1}$ T.~K.~Nelson,$^{1}$ D.~Roberts,$^{1}$
B.~H.~Behrens,$^{2}$
% ...
\end{center}
 
\small
\begin{center}
$^{1}${University of California, Santa Barbara, California 93106}\\
$^{2}${University of Colorado, Boulder, Colorado 80309-0390}\\
% ...
\end{center}

\clearpage

\section{ INTRODUCTION }
\label{s-intro}

The decay~\footnote{Generalization to 
charge conjugate reactions and states is implied throughout, 
except as noted.} 
$\tau^- \to \nu_\tau [3\pi]^-$
has been the subject of much interest over the years.
Because of the transformation properties
of the weak current under parity and G-parity, $\tau$ lepton 
decay to an odd number of pions is expected to occur exclusively through 
the axial vector current, ignoring isospin-violating effects.   
Thus the $3\pi$ system in this decay must have 
spin-parity quantum numbers $J^P = 0^-$ or $1^+$.  As a result of 
this plus the purely weak interaction involved in $\tau$ decay, 
such decays provide an excellent opportunity to investigate the axial vector 
hadronic weak current and the dynamics of axial vector meson decay.  

The $\tau^- \to \nu_\tau [3\pi]^-$ decay  
is dominated by production of the poorly understood 
$a_1(1260)$ meson, 
which is believed to decay mainly via the lowest dimensional 
Born (mostly $S$-wave) $\rho\pi$ intermediate state.  
The world average values~\cite{pdg98} 
for its mass and width are $1230\pm 40$ and 
250--600 MeV respectively, determined primarily from 
$\tau$ decay.  The theoretical understanding of the $a_1$ is not rigorous 
-- many models have been proposed~\cite{bowler,torn,IMR,KS,feindt,ivanov} 
to describe the line shape and resonant substructure, but none have 
provided an entirely satisfactory description of the data.  
Additional experimental input is essential for a better understanding 
of this system.

Recent experimental studies of the $\tau^-\to \nu_\tau \pi^-\pi^+\pi^-$ 
decay have been carried out by the ARGUS~\cite{arg93,ms}, 
OPAL~\cite{opal} and DELPHI~\cite{ronan,delphi} collaborations.  
Analyses of earlier data such as that by Isgur, Morningstar 
and Reader~\cite{IMR} had demonstrated the presence of $D$-wave 
$\rho\pi$ production.  With $\sim 7,500$ events, 
ARGUS~\cite{arg93} measured the ratio of amplitudes
at the nominal $a_1$ mass to be $D/S = -0.11\pm 0.02$.  
In a sample of $\sim 6,300$ events,  
OPAL~\cite{opal} found $D/S = -0.10\pm 0.02\pm 0.02$.  
Another ARGUS analysis~\cite{ms} of $\sim 3,300$ lepton-tagged 
events considered many additional amplitudes.  
This analysis found a signal at the $4.2$ standard deviation 
($\sigma$) level for the presence of an $a_1\to f_2(1270)\pi$ amplitude.  
Neither ARGUS nor OPAL found evidence of 
non-axial-vector contributions such as production of the $J^P=0^-$ 
$\pi^\prime(1300)$ which decays via $\rho\pi$ and $\sigma\pi$.  

In addition, the decays $\tau^-\to \nu_\tau [3\pi]^-$ and 
$\tau^-\to \nu_\tau [5\pi]^-$ have been employed to constrain the 
$\tau$ neutrino mass~\cite{ronan,luca}, 
through investigation of the endpoint in the 
invariant mass and energy spectra of the multi-pion system.  
Notably, ALEPH~\cite{alephnu} has obtained an upper 
limit of 18.2 MeV at the 95$\%$ confidence level (CL) 
on the $\nu_\tau$ mass, based on these decays.  
These analyses rely on an understanding of the 
hadronic dynamics.  DELPHI \cite{ronan,delphi},  
with $\sim 6,500$ events, has reported anomalous structure in 
$\tau^-\to \nu_\tau \pi^-\pi^+\pi^-$ in the context of a 
$\nu_\tau$ mass analysis.  In that work, the Dalitz plot distribution 
for events with very high $3\pi$ mass ($> 1.5\, \mbox{GeV}$) is 
suggestive of enhanced $D$-wave $\rho\pi$ production, while the
$3\pi$ mass spectrum shows an excess in this region relative to 
expectations from a single $a_1$ resonance.  
Inclusion of a radial excitation (an $a_1^\prime$) with a large 
$D$-wave coupling to $\rho \pi$ provides an improved description 
of the DELPHI data, but weakens the $\nu_\tau$ mass limit.  

In this article, we present results from a model-dependent 
analysis of the decay $\tau^-\to \nu_\tau \pi^-\pi^0\pi^0$, based 
on data collected with the CLEO~II detector.  
We perform fits to models to characterize 
both the substructure as seen in Dalitz plot and angular variables, 
as well as the $a_1$ resonance parameters as seen in the $3\pi$ invariant 
mass spectrum.  The $\pi^-\pi^0\pi^0$ channel 
has several advantages relative to the all-charged mode. 
First, the multihadronic and $\tau$ feed-across
backgrounds are smaller.  
Second, the $\pi^-\pi^0\pi^0$ mode may be better suited for discerning 
substructure involving isoscalar mesons because there is only one pairing 
of pions which can have $I=0$, unlike in the all-charged case.  
This second point is particularly relevant 
in light of the ARGUS result for $f_2\pi$ 
production~\cite{ms} and the recent observation of the decay 
$\tau^-\to \nu_\tau f_1(1285)\pi^-$ by CLEO~\cite{vasilii}.  These results 
suggest that isoscalars may play a role in $a_1$ decay.

The outline for the article is as follows.  
In Section~\ref{s-evtsel}, we describe the data sample and event 
selection.  We describe the basic elements of the model used to 
characterize the $\tau^-\to\nu_\tau \pi^-\pi^0\pi^0$ data 
in Section~\ref{s-model}.  In Sections~\ref{ss-substructure}, \ref{ss-nutau} 
and~\ref{ss-massfit}, 
we describe the three analyses of the hadronic structure: (1) performing  
fits to the substructure based on Dalitz plot and angular observables; 
(2) extending these fits to determine the signed $\tau$ neutrino 
helicity; and (3) performing fits to determine the resonant structure 
of the $3\pi$ mass spectrum, making use of results from the 
substructure fits.  We summarize the results and conclude in 
Section~\ref{s-conclude}.

\section{DATA SAMPLE AND EVENT SELECTION}
\label{s-evtsel}

The analysis described here is based on 4.67 fb$^{-1}$ of $e^+e^-$ collision 
data collected at center-of-mass energies  
$2E_{\rm beam}$ of $\sim 10.6$ GeV, 
corresponding to $4.3\times 10^{6}$ reactions of the type 
$e^+e^-\to \tau^+\tau^-$.  These data were recorded at the 
Cornell Electron Storage Ring (CESR) 
with the CLEO~II detector~\cite{cleonim} between 1990 and 1995.  
Charged particle tracking in CLEO~II 
consists of a cylindrical six-layer straw tube array
surrounding a beam pipe of radius 3.2~cm that encloses 
the $e^+e^-$ interaction point (IP), 
followed by two co-axial cylindrical drift chambers of 10 and 51 
sense wire layers respectively.  
Barrel ($|\cos\theta|<0.81$, where $\theta$ is the polar 
angle relative to the beam axis) and end cap scintillation counters 
used for triggering and time-of-flight measurements 
surround the tracking chambers.  For electromagnetic calorimetry, 
7800 CsI(Tl) crystals are arrayed in projective 
(toward the IP) and axial geometries in barrel and end cap sections 
respectively.  The barrel crystals present 16 radiation lengths to 
photons originating from the IP.  

Identification of $\tau^-\to \nu_\tau\pi^-\pi^0\pi^0$ decays relies 
heavily on the segmentation and energy resolution of the calorimeter for 
reconstruction of the $\pi^0$'s.  The central portion of the 
barrel calorimeter ($|\cos\theta|<0.71$) achieves energy and 
angular resolutions of 
$\sigma_E/E\,(\%) = 0.35\,E^{0.75} + 1.9 - 0.1\,E$ and 
$\sigma_\phi\,\mbox{(mrad)} = 2.8/\sqrt{E} + 2.5$, with $E$ in GeV, 
for electromagnetic showers.  The angular resolution ensures that 
the two clusters of energy deposited by the photons from a $\pi^0$ 
decay are resolved over the range of $\pi^0$ energies typical of 
the $\tau$ decay mode studied here.

The detector elements described above are immersed in a 1.5 Tesla 
magnetic field provided by a superconducting solenoid surrounding the 
calorimeter.  Muon identification is accomplished with proportional 
tubes embedded in the flux return steel at depths representing 
3, 5 and 7 interaction lengths of total material penetration 
at normal incidence.

\subsection{Event selection}

To identify events as $\tau\tau$ candidates we require the decay of 
the $\tau^+$ (denoted as the `tagging' decay) that is 
recoiling against our signal $\tau^-$ decay to be classified 
as $\overline{\nu}_\tau e^+\nu_e$, $\overline{\nu}_\tau \mu^+\nu_\mu$, 
$\overline{\nu}_\tau \pi^+$, or $\overline{\nu}_\tau \pi^+\pi^0$.  Thus,  
we select events containing two oppositely charged barrel tracks 
separated in angle by at least $90^\circ$.  To reject backgrounds from 
Bhabha scattering and two-photon interactions we require 
track momenta to be between $0.08\, E_{\rm beam}$ and $0.90\, E_{\rm beam}$. 
Clusters of energy deposition  
in the central region of the calorimeter ($|\cos\theta|<0.71$) 
that are not matched with a charged track projection  
are paired to form $\pi^0$ candidates.  
These showers must have energies greater than 50~MeV, and the 
invariant mass of the photon-pair must lie within $7.5\sigma$ of the 
$\pi^0$ mass where $\sigma$ varies between $\sim 4-7$ MeV.
Those $\pi^0$ candidates with energy above $0.06\, E_{\rm beam}$ 
after application of a $\pi^0$ mass constraint are associated 
with any track within $90^\circ$.  

A $\pi^-\pi^0\pi^0$ candidate is formed from a track which has two 
associated $\pi^0$ candidates as defined above.  
If more than one combination of $\pi^0$ candidates can be assigned 
to a given track, only one combination is chosen: namely, that for 
which the largest unused barrel photon-like cluster in the $\pi^-\pi^0\pi^0$ 
hemisphere has the least energy.  A cluster is defined to be photon-like 
if it satisfies a $1\%$ confidence level cut on the transverse shower 
profile  
and lies at least 30~cm away from 
the nearest track projection.  

As mentioned earlier, to reject background from multihadronic 
($e^+e^- \to q\overline q$) events, the tag system recoiling against 
the $\pi^-\pi^0\pi^0$ candidate must be consistent with $\tau$ 
decay to neutrino(s) plus $e^+$, $\mu^+$, $\pi^+$ or $\pi^+\pi^0$
(denoted as $\rho^+$).  
The recoiling track is identified as an electron if its calorimeter 
energy to track momentum ratio satisfies $0.85<E/p<1.1$ and if its 
specific ionization in the main drift chamber is not less than $2\sigma$ 
below the value expected for electrons.  
It is classified as a muon if the track has penetrated to at least the 
innermost layer of muon chambers at 3 interaction lengths.
If not identified as an $e$ or a $\mu$, then if the track is accompanied by 
a third $\pi^0$ of energy $\ge 350\,{\mbox MeV}$ which lies closer
to it than to the $\pi2\pi^0$ system, the track-$\pi^0$ combination 
is classified as a $\rho$ tag.  The invariant mass of this system must
be between 0.55 and 1.20 GeV.  If not identified as an 
$e$, $\mu$, or $\rho$ tag, the recoil track is identified as a single 
$\pi$ tag.  
To ensure that these classifications are consistent with 
expectations from $\tau$ decay, events are vetoed if any unused 
photon-like cluster with $|\cos\theta| < 0.95$ 
has energy greater than 200 MeV, or if any 
unmatched non-photon-like cluster has energy above 500 MeV.  
The missing momentum as determined using the $\pi2\pi^0$ and tagging 
systems must point into a high-acceptance region of the detector 
($|\cos\theta_{\rm miss}| < 0.9$), and must have a component 
transverse to the beam of at least $0.06\, E_{\rm beam}$.

\subsection{Final event sample}
\label{ss-evtsample}

\begin{figure}
  \centering \leavevmode
  \epsfysize=7.in
  \epsfbox{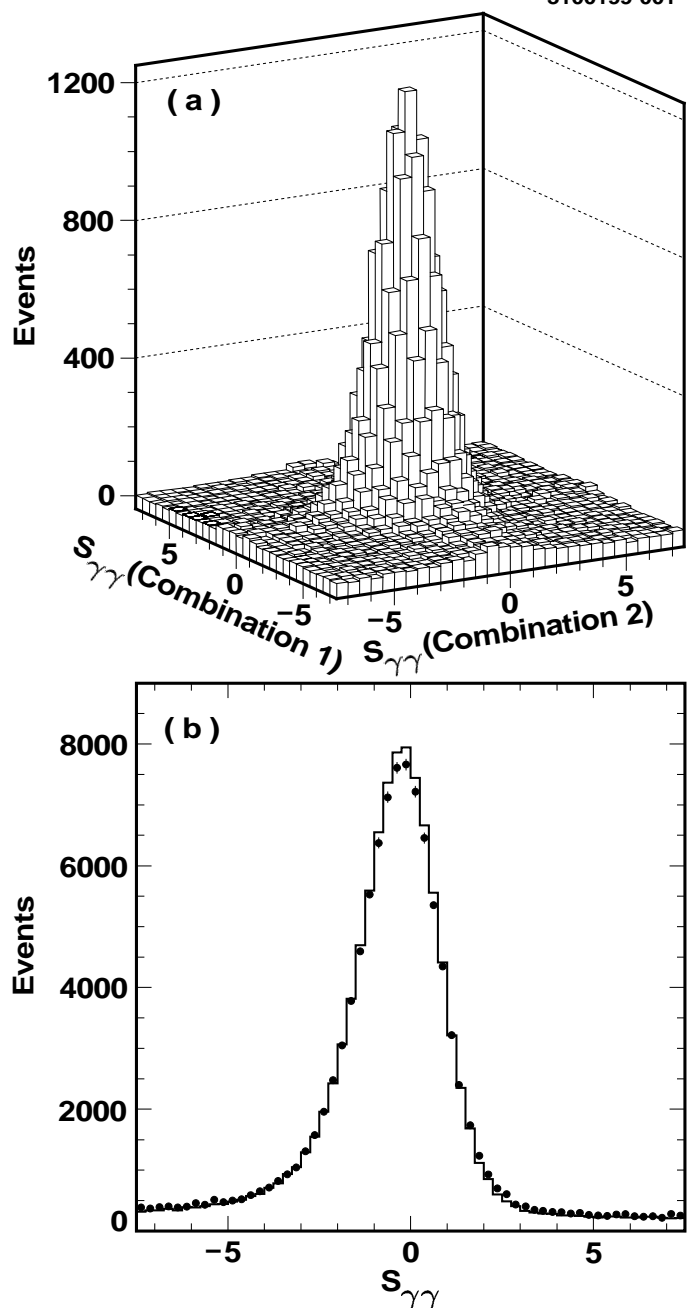}
  \caption[]{\small (a) Plot of the normalized photon-pair invariant
             mass $S_{\gamma\gamma} = (M_{\gamma\gamma}-m_{\pi^0}) 
             /\sigma_{\gamma\gamma}$ for the two $\pi^0$ candidates 
             after all other cuts have been applied. 
             (b) Comparison of the $S_{\gamma\gamma}$
             distribution (two entries per event) for data (points) and 
             $\tau\tau$ Monte Carlo (histogram) samples.
            }
  \label{fig:sggvsgg}
\end{figure}

After all cuts, the remaining sample consists of 51,136 events.  
The normalized invariant masses of the two photon-pairs,  
$S_{\gamma\gamma} \equiv (M_{\gamma\gamma}-m_{\pi^0})/\sigma_{\gamma\gamma}$,  
are plotted
against each other in Fig.~\ref{fig:sggvsgg}(a).  
In Fig.~\ref{fig:sggvsgg}(b), $S_{\gamma\gamma}$ is 
plotted for all $\pi^0$ candidates along with the corresponding 
distribution from the $\tau$ Monte Carlo (MC) sample described in 
the following section.
We define the $\pi^0\pi^0$ signal region to be that where 
$-3.0 < S_{\gamma\gamma}< 2.0$ for both $\pi^0$ candidates.  
In the $\pi^0\pi^0$ signal region there are 36710 events, of which 
17234 are tagged by leptonic decays of the recoiling $\tau$.  
To estimate the contributions from fake $\pi^0$'s, we also define side 
and corner band regions using 
$-7.5< S_{\gamma\gamma} < -5.0$ and $3.0 < S_{\gamma\gamma} < 5.5$. 

\begin{figure}
  \centering\leavevmode
  \epsfxsize=6.in
  \epsfbox{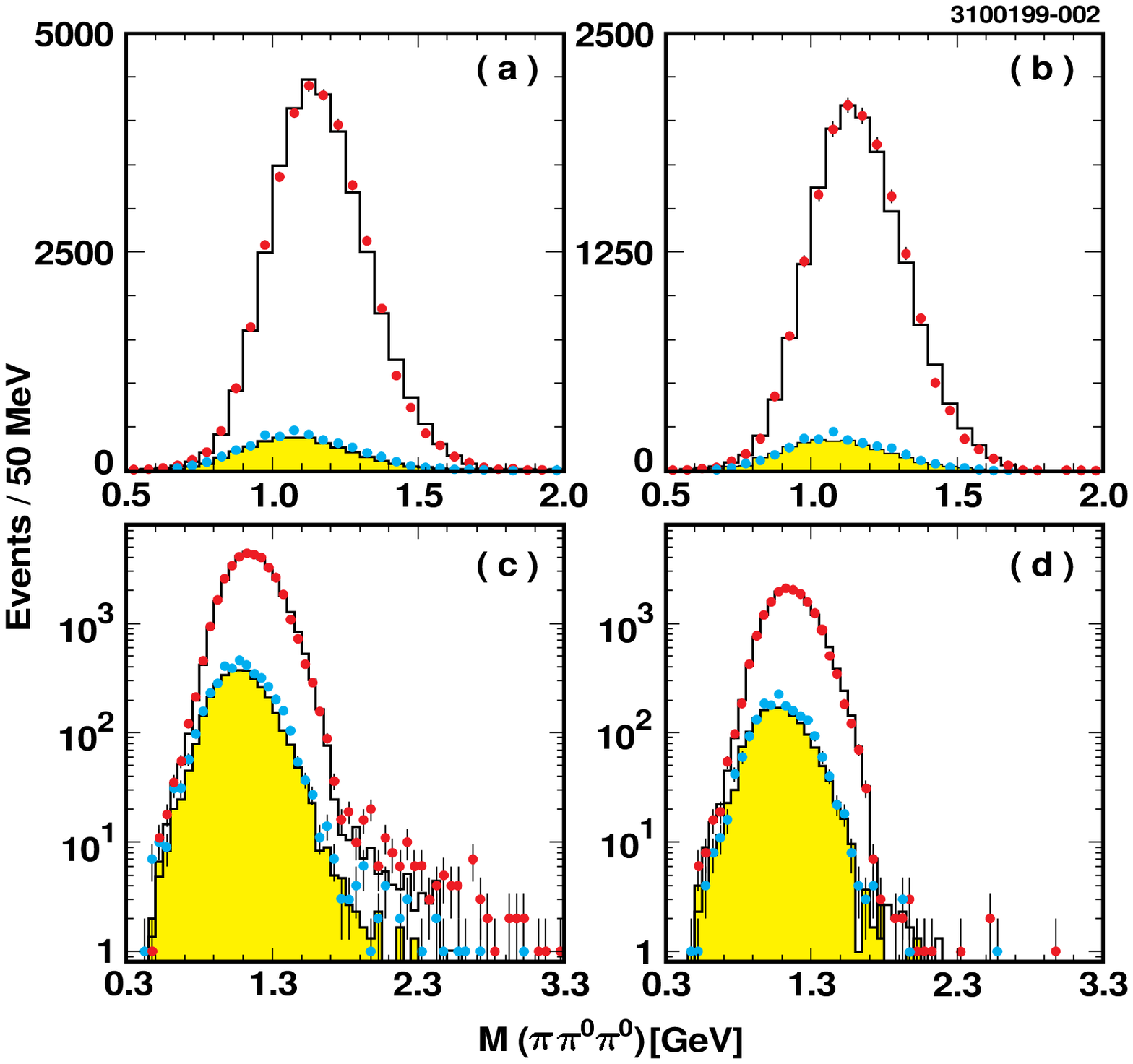}
  \caption[]{\small 
             $\pi^-\pi^0\pi^0$ mass spectrum from the all-tag (a) 
             and lepton tag (b) samples, after cuts.   Shown are 
             events in the $\pi^0\pi^0$ signal (dark points for the 
             data, unshaded histogram for the Monte Carlo spectrum) and 
             side band (light points for the data, shaded histogram 
             for the MC spectrum) regions.      
             Plots (c) and (d) show the same spectra as in (a) and (b)
             plotted over a larger range and on a logarithmic scale.
            }
  \label{fig:m3pi}
\end{figure}

In Fig.~\ref{fig:m3pi}, we plot the $\pi^-\pi^0\pi^0$ mass for events
in the $\pi^0\pi^0$ signal and side band regions, 
for data and $\tau$ MC samples.
The events above the $\tau$ mass 
in Fig.~\ref{fig:m3pi}(c) are dominantly due to 
feedacross from $\tau\to \rho \nu$ decays 
where the second $\pi^0$ is being picked up from the recoil $\tau$ decay.  
The $\tau$ Monte Carlo simulation 
accounts for most of this high mass tail, but 
not all, with the remainder being due to a small 
$q\overline q$ background contribution. 
The high mass events are essentially absent from the lepton-tagged 
events, plotted in Fig.~\ref{fig:m3pi}(d).

\subsection{Monte Carlo samples}
\label{ss-mc}

For determination of detection efficiency for our signal decay as well as 
for backgrounds from other $\tau$ decay modes, we rely on a sample of 
Monte Carlo $e^+e^-\to \tau^+\tau^-$ events with equivalent luminosity 
approximately three times that of the data.  These events were generated 
using the {\tt KORALB/TAUOLA}~\cite{korb} program, and then passed through 
the {\tt GEANT}-based~\cite{geant} CLEO~II detector simulation package.  
The full CLEO event reconstruction program was then run on this sample. 
The MC distributions shown in Figs.~\ref{fig:sggvsgg} 
and~\ref{fig:m3pi} are derived from this sample.  

In {\tt TAUOLA}, the $\tau^-\to \nu_\tau [3\pi]^-$ decay is described 
with a single $a_1$ resonance decaying solely via the lowest dimensional 
($s$-wave, in the notation introduced in the next section) 
Born amplitude for $(\rho +\rho^\prime)\, \pi$ production, 
following the model of K\"uhn and Santamaria~\cite{KS}.    
We have tuned the $a_1$ mass and width 
to yield a $3\pi$ mass spectrum that roughly matches that seen in our 
data in the all-charged mode.  Although the data and MC 
$3\pi$ mass spectra show reasonable agreement on average 
(see Fig.~\ref{fig:m3pi}), close inspection reveals significant 
deviations, particularly in the high mass region located roughly between 
1.4 GeV and the $\tau$ mass.  The Dalitz plot 
distributions agree poorly with the corresponding MC distributions, 
especially in the $M_{\pi^0\pi^0}$ 
projection and most strikingly at high $3\pi$ mass.

For the substructure fits described in Section~\ref{ss-substructure}, we 
generated additional MC samples for our signal mode plus key 
background $\tau$ decay modes.  For these samples,   
we developed special purpose event generators. 
Unlike the treatment in {\tt KORALB/TAUOLA}, we implemented 
radiative effects according to an approximation in which they factorize 
from the rest of the differential matrix element, as required by the 
reverse Monte Carlo approach described in Section~\ref{sss-fitmeth}.  
In addition, the signal mode was generated with a $3\pi$ mass spectrum 
weighted towards high values so as to ensure high statistics in 
the high-mass region where the data show the most apparent deviation 
from the model used by {\tt TAUOLA} in both Dalitz plot and $3\pi$ 
mass distributions.

\section{MODEL OF \boldmath $\tau^- \to \nu_\tau \pi^-\pi^0\pi^0$ }
\label{s-model}

Tau lepton decay to neutrino plus three pions follows the form 
\begin{equation}
\label{eq:taudec}
d\Gamma(\tau \to \nu_\tau 3\pi) = \frac{1}{2m_\tau} 
   \left[ \frac{G_F^2 V_{ud}}{2} L^{\mu\nu} J_\mu J^*_\nu\right] d\Phi_4,
\end{equation}
where $L^{\mu\nu}$ represents the lepton tensor for weak decay, $J^\mu$ 
denotes the hadronic weak current for production of three pions, and 
$d\Phi_4$ is the Lorentz-invariant 4-body phase space element for the decay.
The goal of this analysis is to probe the structure of the hadronic 
current, benefitting from the well-understood properties of the 
weak interaction.  

In principle $J^\mu$ is comprised of vector and 
axial vector currents:
\begin{equation}
\label{eq:current}
 J^\mu = J_V^\mu + J_A^\mu = < 3\pi | \overline{d}\gamma_\mu u | 0 >
                           + < 3\pi | \overline{d}\gamma_5\gamma_\mu u | 0 > , 
\end{equation}
however $G$-parity conservation requires that $J^V_\mu = 0$.  Thus 
we consider only contributions from the axial vector current.

In $\tau$ decay, the squared momentum transfer $s\equiv M_{3\pi}^2$ is 
small, and thus the dynamics are expected to be dominated by resonance 
production. 
The possible axial vector ($J^P=1^+$) 
resonance contributions are the $a_1(1260)$ and 
radial excitations, {\sl i.e.}, the $a_1^\prime$.  In addition, pseudoscalar 
($J^P=0^-$) contributions are possible, {\sl i.e.}, the $\pi^\prime(1300)$, 
although these are expected to be suppressed according to the 
Partially Conserved Axial Current (PCAC) hypothesis.  
In this section we describe the model used to parametrize 
$J^\mu$, assuming the $3\pi$ system is in a $J^P=1^+$ state. 

\subsection{Model for substructure in \boldmath $1^+ \to 3\pi$}
\label{ss-modsub}

The strong decay of the $a_1$ is expected to involve 
substructure which is again dominated by resonance production.  
We write for the contribution to $J^\mu$ involving $a_1$ production
\begin{equation}
\label{eq:jmu}
  J^\mu_{a_1} = B_{a_1}(s) \times \sum_i \beta_i \, j^\mu_i , 
\end{equation}
where $B_{a_1}(s)$ denotes the $a_1$ Breit-Wigner, $\beta_i$ are complex 
coupling constants, and $j^\mu_i$ contain form factors describing 
components of the substructure involving specific resonances.  
The details of this parametrization are given in 
Appendix~\ref{a-substructure}.  
As an example, in the case of $s$-wave $\rho\pi$ production, 
\begin{equation}
\label{eq:rhopi}
  j^\mu_{\rho\pi,\, s\mbox{-wave}} = T^{\mu\nu} \left[
         q_{1\nu}\, B_\rho(s_1)\, F_{R_{\rho\pi}}(k_1) 
       - q_{2\nu}\, B_\rho(s_2)\, F_{R_{\rho\pi}}(k_2) \right].   
\end{equation}
We define $p_1$, $p_2$ and $p_3$ as the four-momenta of the three pions, 
in our case $\pi^0_1$, $\pi^0_2$ and $\pi^-$, respectively.  We define 
$a = p_1 + p_2 + p_3$, $s_i = (p_j + p_k)^2$ 
and $q_i = p_j - p_k$, where $(i,j,k)$ 
represent cyclic permutations of $(1,2,3)$.   
The factor $T^{\mu\nu}$ denotes the expression 
$g^{\mu\nu} - a^\mu a^\nu/a^2$.  
The factors $B_\rho(s_i)$ 
denote Breit-Wigners describing the corresponding $\rho \to \pi\pi^0$ 
amplitudes.   Finally, we have included an additional form factor 
$F_{R_{\rho\pi}}$, which represents the effect of the finite size of the 
$a_1$ meson in its decay to $\rho\pi$.  We take this form factor to 
have the form 
\begin{equation}
\label{eq:mesonradius}
  F_{R_i}(k_i) = e^{-R_i^2 k_i^2/2}, 
\end{equation} 
where 
$k_i$ is the momentum of the decay products, the $\rho$ and 
the $\pi$ in this case, in the $a_1$ rest frame.
The parameter $R_i$ is proportional 
(by a factor of $\sqrt{6}\hbar c$, see Ref.~\cite{torn}) 
to the root mean square (r.m.s.) radius of the $a_1$.  
We note that expressions for $j^\mu_i$ must be 
symmetric with respect to interchange of $\pi^0_1$ and $\pi^0_2$ 
since these are indistinguishable. 

In our analysis of substructure in the $\tau\to\nu_\tau 3\pi$ decay, 
we consider the following amplitudes:  
\begin{equation}
  \label{eq:amplist}
  \begin{array}{cl}
j_1^\mu : & \mbox{$s$-wave amplitude for $1^+ \rightarrow \rho(770)\pi$  
    (denoted as $\rho\pi$)}, \\
j_2^\mu : & \mbox{$s$-wave amplitude for $1^+ \rightarrow \rho(1450)\pi$  
    (denoted as $\rho^\prime\pi$)}, \\
j_3^\mu : & \mbox{$d$-wave amplitude for $1^+ \rightarrow \rho(770)\pi$}, \\
j_4^\mu : & \mbox{$d$-wave amplitude for $1^+ \rightarrow \rho(1450)\pi$}, \\
j_5^\mu : & \mbox{$p$-wave amplitude for $1^+ \rightarrow f_2(1270)\pi$}, \\
j_6^\mu : & \mbox{$p$-wave amplitude for $1^+ \rightarrow f_0(400-1200)\pi$   
   (denoted as $\sigma\pi$)}, \\ 
j_7^\mu : & \mbox{$p$-wave amplitude for $1^+ \rightarrow f_0(1370)\pi$}.
  \end{array}
\end{equation}
An explicit parametrization of the amplitudes $j_i^\mu$ is given 
by Eqn.~\ref{eq:2pi0pi}.  
With these definitions, the constants $\beta_i$ have dimensions 
of [GeV]$^x$, where the exponent depends on the amplitude.  
In our fits we specify $\beta_1=1$, 
such that the couplings for the other amplitudes are determined relative 
to the $s$-wave $\rho\pi$ coupling.  The parameters used to describe the 
resonances appearing in the $j^\mu_i$ above are given in 
Table~\ref{tab:res_par}, while the Breit-Wigner form used here is given in 
Appendix~\ref{a-substructure} by Eqn.~\ref{eq:bwdef}.  
\begin{table}
\begin{center}
\caption[]{\small Resonance parameters for the intermediate states 
                  as used in the substructure fits.}
\label{tab:res_par}
\begin{tabular}{lccc}
\multicolumn{1}{c}{$Y$} & $m_{0Y}$ & $\Gamma_0^Y$ & Ref. \\ 
              & [GeV] & [GeV] & \\ \hline
$\rho(770) $  & $0.774$ & $0.149$ & \cite{ju}\\
$\rho(1450)$  & $1.370$ & $0.386$ & \cite{ju}\\
$f_2 (1270)$  & $1.275$ & $0.185$ & \cite{pdg98}\\
$\sigma    $  & $0.860$ & $0.880$ & \cite{tornscal}\\
$f_0 (1370)$  & $1.186$ & $0.350$ & \cite{tornscal}\\ 
\end{tabular}
\end{center}
\end{table}

In Eqns.~\ref{eq:rhopi} and~\ref{eq:2pi0pi}, 
we have constructed Lorentz-invariant amplitudes so as to make 
contact with the resonant components of the substructure.    
In contrast with a formulation based on angular momentum eigen 
functions, these amplitudes are only approximately associated with a 
specific angular momentum quantum number $L$, and hence we have employed 
lower case letters to identify the primary value of $L$.  Thus, for 
example the lowest dimensional Born amplitude for $\rho\pi$, the 
Lorentz-invariant $s$-wave amplitude, contains a small $D$-wave component 
(see for example, Refs.~\cite{IMR,feindt}). 

The selection of the amplitudes $j^\mu_1,\ldots,j^\mu_7$ is in part 
based on experience gained in early attempts to fit the data.  It is 
also in part motivated by the unitarized quark model 
of T\"ornqvist~\cite{tornscal}. 
The resonance parameters of the broader $\sigma$ and $f_0(1370)$ mesons are 
taken from application of this model to existing data~\cite{tornscal}.  
We have also performed fits with additional amplitudes, namely the
axial vector $f_0(980)\pi$ and pseudoscalar $\pi^\prime \to \rho\pi$ and 
$\sigma\pi$.  These are discussed in Section~\ref{sss-submodvar}.

\subsection{Model for the \boldmath $3\pi$ mass spectrum}
\label{ss-modmass}

The conventional understanding of 
$\tau\to\nu_\tau\, 3\pi$ decay is that it proceeds through creation 
of the lowest lying axial vector meson, the $a_1(1260)$.  
Since radial excitations may also be present,   
we replace the Breit-Wigner function $B_{a_1}(s)$ 
appearing in Eqn.~\ref{eq:jmu} by a modified function
that includes a possible $a_1^\prime$ admixture. 
\begin{eqnarray}
\label{eq:breit}
B(s)= B_{a_1} (s) + \kappa\cdot B_{a^\prime_1} (s)
&=& \frac{1}{s-m^2_{a_1}(s)+im_{0\,a_1}\Gamma^{a_1}_{tot} (s)} \nonumber\\  
&+&\frac{\kappa}{s-m^2_{0\, a_1^\prime} +im_{0\,a_1^\prime} \,
\Gamma^{a_1^\prime}_{tot}(s)} \, , 
\end{eqnarray}
where $\kappa$ is an unknown complex coefficient.
The $a_1^\prime$ meson is predicted in a flux-tube-breaking model
\cite{KI,GI} with a mass of 
$m_{0\, a_1^\prime} = 1.820 \mbox{ GeV}$.  Experimental 
indications~\cite{ves} suggest a mass of $1.7 \mbox{ GeV}$ and 
a width of $0.3$ GeV.  One impact of introducing the $a_1^\prime$ 
in this way is that the coupling constants $\beta_i$ in Eqn.~\ref{eq:jmu} 
will necessarily vary with $s$.  We will return to this issue 
later in this article.

In Eqn.~\ref{eq:breit}, the function $m^2_{a_1} (s)$ is the 
running mass \cite{IMR,tornscal}, 
\begin{equation}
\label{eq:runmass}
m^2_{a_1} (s) = m_{0\, a_1}^2 + \delta^2 (s) \, ,
\end{equation}
where $\delta^2 (s) $ is the mass shift function, 
\begin{equation}
\label{eq:disper}
\delta^2 (s) = \frac{1}{\pi} \int_{s_{\rm min}}^\infty 
\frac{ m_{0\, a_1} \Gamma^{a_1}_{tot} (s^\prime )}{s- s^\prime} ds^\prime \, . 
\end{equation}
The mass shift function is renormalized such that
\begin{equation} 
m_{a_1} (s)\vert_{s = m^2_{0\, a_1}} = m_{0\, a_1} \, . 
\end{equation}
The bare mass $m_{0\, a_1}$ is chosen to be the resonance mass by 
requiring that the total width $\Gamma^{a_1}_{tot} (s)$ at $s=m^2_{0\, a_1}$ 
is equal to the nominal width $\Gamma_{0\, a_1}$.

The $\sqrt{s}$-dependent behavior of the $a_1$ width, and consequently 
of its mass, requires knowledge of the underlying substructure, not 
just for $a_1 \to 3\pi$, but also 
for decays to other channels such as $a_1\to K\overline{K}\pi$ 
[via $a_1 \to K^*K$ and $a_1 \to f_0(980)\pi$]. 
Considering only these contributions, the $a_1$ width can be written as 
\begin{equation}
\label{eq:runwidth}
\Gamma_{tot}^{a_1}(s)= g^2_{a_1 (3\pi)}
\left(
                                    \hat{\Gamma}_{2\pi^0\pi^\mp}^{a_1}   (s) + 
                                    \hat{\Gamma}_{2\pi^\mp\pi^\pm}^{a_1} (s) +
      \gamma^2_{a_1 (K^\star K)}    \hat{\Gamma}_{K^\star K}^{a_1}       (s) +
      \gamma^2_{a_1 (f_0 (980) \pi)}\hat{\Gamma}_{f_0 (980) \pi}^{a_1}   (s) 
\right) \, ,
\end{equation}
where $g_{a_1 (3\pi)}$ denotes the coupling of the $a_1$ meson to the 
$3\pi$ system, the $\gamma_{a_1 (x)}$ denote the relative coupling of $x$ to 
the $a_1$ meson, and the $\hat{\Gamma}^{a_1}_{x}$ denote the reduced widths.

The $a_1 \to 3\pi$ partial width can be expressed in terms 
of the amplitudes for the hadronic current  $j^\mu_i$, 
as defined by Eqn.~\ref{eq:jmu}.   Specifically, the reduced widths  
$\hat{\Gamma}^{a_1}_{2\pi^0\pi^\mp}$ and 
$\hat{\Gamma}^{a_1}_{2\pi^\mp\pi^\pm}$ are
\begin{equation}
\label{eq:a13pi}
\hat{\Gamma}^{a_1}_{2\pi^0\pi^\mp/2\pi^\mp\pi^\pm } = \int \sum_{ij} \left[ 
- \beta_i  \beta_j^\star \, j_{i\mu} \, j_j^{\star\mu}       \right]
d\Phi_{3\pi} \, ,
\end{equation}
where $d\Phi_{3\pi}$ denotes 3-body phase space for the 
$a_1^- \to \pi^-\pi^0\pi^0$ decay.  
We determine $\hat{\Gamma}^{a_1}_{2\pi^0\pi^\mp}$ numerically 
using the output from the substructure analysis in which the 
$j_i^\mu$ are specified according to Eqn.~\ref{eq:2pi0pi} 
in Appendix~\ref{a-substructure}.  
Similarly, for $\hat{\Gamma}^{a_1}_{2\pi^\mp\pi^\pm}$ we make use of 
isospin symmetry to infer the corresponding all-charged amplitudes from 
our analysis of the $\pi^-\pi^0\pi^0$ substructure. 

The $K^*K$ and $f_0(980)\pi$ partial widths contribute as thresholds in 
the $\sqrt{s}$-dependence of the $a_1$ width.  We determined these 
from expressions for the $s$- and $p$-wave amplitudes, respectively, 
making use of the narrow width approximation for the 
$f_0$ and $K^*$.  The relative couplings for these contributions 
are left as free parameters to 
be determined from the data, along with the $a_1$ pole mass and width.

\section{ANALYSIS OF DALITZ PLOT AND ANGULAR VARIABLES}
\label{ss-substructure}

The primary goal of this analysis is to characterize the 
contributions to the substructure of the $a_1\to 3\pi$ decay,  
as well as the parameters describing $a_1$ line shape itself, 
including the question of possible radial excitations.  
Two separate analyses are carried out to address these two issues, 
however it is important to realize that they are closely coupled.  

First, the integration over the Dalitz plot needed to specify the 
mass-dependence of the $a_1$ width as well as the running of the 
$a_1$ mass requires the amplitudes participating in the 
$3\pi$ hadronic current to have been determined.  
On the other hand, the question of whether an $a_1^\prime$ resonance also  
contributes to the $3\pi$ mass spectrum affects the way one would choose 
to parametrize the substructure.

Practically, it is not feasible to fit the 
$3\pi$ mass spectrum and the hadronic substructure simultaneously.
We have elected to determine first the substructure in a way that is 
mostly independent of the $3\pi$ mass spectrum.   Then in a second step, 
using the results on $\hat{\Gamma}^{a_1}_{3\pi}(s)$ 
obtained in the substructure fits,
we measure the $a_1$ resonance parameters from the $3\pi$ mass spectrum.  
In this section we describe the substructure fits, while the fits to 
the $3\pi$ mass spectrum are described in Section~\ref{ss-massfit}.  


\subsection{Fitting method}
\label{sss-fitmeth}

To determine the contributions to the substructure in 
$\tau\to \nu_\tau 3\pi$, we perform unbinned maximum 
likelihood fits using as input 
the measured 3-momenta of the three pions in the decay,  
and the energy of the decaying $\tau$ lepton.  
The latter is known to be the beam energy 
in the absence of radiative effects.   
With knowledge of the particle masses (we take the 
mass of the $\tau$ neutrino to be zero),   
these inputs comprise a complete kinematical description of the decay, 
with the exception of: (1) deviations in the $\tau$ energy due to initial 
state radiation (ISR), (2) the azimuthal orientation $\varphi_\tau^{3\pi}$
of the $\tau$ flight direction  
relative to the measured momentum vector of the $3\pi$ system, 
and (3) smearing due to scattering and detector resolution, 
the beam energy spread, and radiative effects other than ISR.

Following the discussions in Sec.~\ref{s-model} and 
Appendix~\ref{a-substructure}, and 
ignoring the sources of smearing described in item (3) above, 
we construct the likelihood function.  The numerator $d\sigma(p_1,p_2,p_3)$ 
of the likelihood ${\cal L}=d\sigma / \int d\sigma$ is
\begin{eqnarray}
\label{eq:likel}
d\sigma 
        & = & \left[ \int \left\{  
  P\left(\cos\theta^{lab}_{3\pi},
         \cos\theta_{\tau}^{3\pi}(E_{3\pi}^{lab},|p_\gamma|),
         \varphi_{\tau}^{3\pi} \right)
              \times f(p_\gamma) 
              \times \left( S_{\mu\nu} + i h_{\nu_\tau} A_{\mu\nu} \right) 
                          \right\} 
              d^3{p_\gamma} d\varphi_{\tau}^{3\pi} \right] \nonumber \\
        & & 
\times \left[ \sum_{ij} ( \beta_i \, j^\mu_i ) 
       ( \beta^\star_j \, j^{\star\nu}_j ) 
\times {\vert B (s) \vert }^2 \right] 
d\!\cos{\theta_{3\pi}^{lab}}\,d\varphi_{3\pi}^{lab}\,
dE^{lab}_{3\pi}\,d\Phi_{3\pi}\,ds\, ,  
\end{eqnarray}
where we integrate over the unmeasured $\nu_\tau$ 
($\varphi_\tau^{3\pi}$ above) and ISR photon ($p_\gamma$) 
degrees of freedom, such that $d\sigma$ is a function of the 
measured degrees of freedom.  For illustrative purposes we 
represent these in the above by the squared $3\pi$ invariant mass ($s$),  
the energy and orientation of the $3\pi$ system in the laboratory
($E_{3\pi}^{lab},\,\theta_{3\pi}^{lab},\,\varphi_{3\pi}^{lab}$), 
and the 3-body phase space element ($d\Phi_{3\pi}$).  
The phase space factor can be expressed in terms of 
the Dalitz plot variables $s_1 = M_{\pi^-\pi^0_2}^2$ 
and $s_2 = M_{\pi^-\pi^0_1}^2$ and the Euler angles 
representing the orientation of the $3\pi$ 
decay plane in the $3\pi$ rest frame.  
The symbols $S_{\mu\nu}$ and $A_{\mu\nu}$ represent the 
symmetric and antisymmetric terms in the lepton tensor.
The factor $f(p_\gamma)$ denotes the factorized ISR photon probability 
distribution.  Finally, we have also included the 
$\tau$-pair production dynamics, the effect of which is to make 
non-uniform the probability distribution, denoted by the factor $P$, 
for the azimuthal angle $\varphi_{\tau}^{3\pi}$.   
The polar angle $\theta_{\tau}^{3\pi}$ between the  
$\tau$ direction and the $3\pi$ system appearing in this factor 
is determined by $E_{\rm beam}$, $E^{lab}_{3\pi}$ and 
$|p_\gamma|$.  
The $\tau$ neutrino helicity $h_{\nu_\tau}$ and the complex coupling
constants $\beta_i$ of the hadronic amplitudes are the fit parameters.

The above integral is computed using a reverse Monte Carlo 
technique~\cite{ms,mscleo}.  In this method, for each event in the data 
we generate a sample of trial MC events which are designed to have 
precisely the measured values for the pion momenta, but which have 
unmeasured quantities determined randomly according to 
the factorized distributions for ISR photons 
and the unknown azimuthal angle $\varphi_{\tau}^{3\pi}$.  
The integration is performed using trial events that possess 
internally consistent kinematics. 
We remove data events for which the number $n_{hit}$ of these 
successful trials is low so as to maintain high precision 
on the integration.  This requirement also tends to 
preferentially remove background events.

To be insensitive to details of the $3\pi$ mass spectrum 
(addressed in Section~\ref{ss-massfit}), we subdivide 
the data in fine bins (25 MeV) of $\sqrt{s}$ and calculate the 
normalization $N=\int d\sigma$ of the likelihood separately 
for each bin $j$: 
\begin{equation}
N_j = \int_{s_j}^{s_{j+1}} \left[ \int 
\frac{\epsilon\cdot d\sigma}
  {d\!\cos{\theta_{3\pi}^{lab}}\,d\varphi_{3\pi}^{lab}\,
   dE^{lab}_{3\pi}\,d\Phi_{3\pi}\,ds}
\, d\!\cos{\theta_{3\pi}^{lab}}\,d\varphi_{3\pi}^{lab}\,
   dE^{lab}_{3\pi}\,d\Phi_{3\pi}\right] \,ds \, ,
\end{equation}
where $\epsilon$ denotes the detector efficiency.  Over the bin width 
$\Delta s = s_{j+1} - s_j$,  ${\vert B (s) \vert }^2$ 
is approximated to be constant, and thus cancels in the likelihood.  
The normalization integrals are 
computed using factorization-based Monte Carlo events 
that have been passed through the 
full detector simulation as described in Section~\ref{ss-mc}.

\subsection{Treatment of backgrounds}

In addition to the likelihood for signal events defined by 
Eqn.~\ref{eq:likel}, we also include the four main background sources 
listed in Table~\ref{tab:back_m3pi}.  There, the background fractions,  
estimated from the $\tau$ MC sample 
for the $\nu 4\pi$, $\nu K\pi\pi$ and $\nu Ks \pi$ modes, 
are tabulated in slices of $\sqrt{s}$ so as to illustrate the dependence.
\begin{table}
\begin{center}
\caption[]{\small Background contributions ($\%$) to the lepton-tagged
sample in slices of the $3\pi$ invariant mass $\sqrt{s}$ 
after the cuts on $\sqrt{s} > 0.6 \mbox{ GeV}$ and $n_{hit}$.}
\label{tab:back_m3pi}
\begin{tabular}{lcccc}
                                                        & 
fake $\pi^0$                                            &  
$\tau^-\rightarrow\nu_\tau \pi^- 3\pi^0 $               & 
$\tau^-\rightarrow\nu_\tau K^-\pi^0\pi^0$               & 
$\tau^-\rightarrow\nu_\tau K_S\pi^- $
\\ \hline
Bin 1\phantom{-8}: 0.6-0.9 $\mbox{ GeV}$&
$  17.2 \pm 1.5    $ & $  14.0 \pm 1.6    $ & $  2.1 \pm 0.6    $ & 
$   8.7 \pm 1.3    $ \\ 
Bin 2\phantom{-8}: 0.9-1.0 $\mbox{ GeV}$&
$  12.9 \pm 0.8    $ & $   5.5 \pm 0.6    $ & $  0.7 \pm 0.2    $ & 
$   1.5 \pm 0.3   $ \\ 
Bin 3\phantom{-8}: 1.0-1.1 $\mbox{ GeV}$&
$   9.5 \pm 0.5    $ & $   4.3 \pm 0.4    $ & $  0.5 \pm 0.1    $ & 
$   0.1 \pm 0.1    $ \\ 
Bin 4\phantom{-8}: 1.1-1.2 $\mbox{ GeV}$&
$   6.8 \pm 0.4    $ & $   3.0 \pm 0.3    $ & $  0.6 \pm 0.1    $ & 
$   0.1 \pm 0.1    $ \\ 
Bin 5\phantom{-8}: 1.2-1.3 $\mbox{ GeV}$&
$   7.1 \pm 0.5    $ & $   2.3 \pm 0.3    $ & $  0.7 \pm 0.2    $ & 
$    0.0    $ \\ 
Bin 6\phantom{-8}: 1.3-1.4 $\mbox{ GeV}$&
$   6.5 \pm 0.6    $ & $   1.8 \pm 0.3    $ & $  0.4 \pm 0.2    $ & 
$    0.0    $ \\ 
Bin 7\phantom{-8}: 1.4-1.5 $\mbox{ GeV}$&
$   6.6 \pm 0.9    $ & $   0.9 \pm 0.3    $ & $  0.2 \pm 0.2    $ & 
$    0.0    $ \\ 
Bin 8\phantom{-8}: 1.5-1.8 $\mbox{ GeV}$&
$   6.8 \pm 1.3    $ & $   0.3 \pm 0.3    $ & $  0.2 \pm 0.2    $ & 
$    0.0    $ \\ 
\hline 
Bin 1-8:           0.6-1.8 $\mbox{ GeV}$&
$   8.5 \pm 0.2    $ & $   3.5 \pm 0.2    $ & $  0.6 \pm 0.1    $ & 
$   0.5 \pm 0.1    $ \\ 
\end{tabular}
\end{center}
\end{table}

Events with fake $\pi^0$'s tend to be $\tau^-\to\nu_\tau\rho^-$ events 
where a spurious $\pi^0$ has been recontructed from clusters associated 
with radiative photons, shower fragments from the interaction of the 
charged $\pi$ in the detector, or other accidental activity in the 
calorimeter.
The likelihood distribution for the fake $\pi^0$ background is 
approximated from data by the Dalitz plot distribution of 
events populating the $\pi^0$ mass side bands.  

For the $\tau^-\rightarrow\nu_\tau\pi^- 3\pi^0$ background 
the reverse Monte Carlo procedure 
is modified to simulate a lost $\pi^0$.  
The $4\pi$ matrix element is not well measured.
We consider models in which the $4\pi$ system 
arises via the $\rho(1450)$ resonance, 
where we simulate either (1) $\rho^{\prime -}\to a_1^-\pi^0\;(s\mbox{-wave})
\to \rho^-\pi^0\pi^0$, 
or (2) $\rho^{\prime -}\to \rho^-\sigma\; (s\mbox{-wave})$, 
or a combination thereof.
%
The Dalitz plot projections from these models are very
similar.  In addition, the goodness of fit varies little 
with the choice of model. 
In the fits reported here, we used 
the $\rho^{\prime -}\to \rho^-\sigma$ model.   

The background $\tau^-\rightarrow \nu_\tau K^-\pi^0\pi^0$ is modeled
by the decay chain
$ 
\tau^-\rightarrow \nu_\tau K_1^- \mbox{, }  
K_1^-\rightarrow K^{\star -} \pi^0 \mbox{ ($s$-wave), }
$ 
where the $K_1$ meson is parametrized by
a superposition of the $K_1 (1270)$ and $K_1(1400)$ Breit-Wigner 
functions.  
Finally, 
the $\tau^-\rightarrow\nu_\tau K_S\pi^- $ background is parametrized by
the decay chain $\tau^- \rightarrow \nu_\tau K^{\star -}$, 
$K^{\star -} \rightarrow K_S^0 \pi^-$ ($p$-wave). 
The mass distribution for the $K_S\to\pi^0\pi^0$ decay 
is parametrized by a Gaussian, 
where the mean and the width are taken from data.

With the inclusion of these backgrounds the likelihood function is:
\begin{eqnarray}
{\cal L} & = &
( 1 - \alpha_{f_{\pi^0}} - \alpha_{4\pi} 
    - \alpha_{K\pi\pi} -\alpha_{K_S \pi} ) 
                   {\cal L}_{signal}  \nonumber \\
& & \mbox{} +
\alpha_{f_{\pi^0}} {\cal L}_{f_{\pi^0}} +
\alpha_{4\pi}      {\cal L}_{4\pi}     +
\alpha_{K\pi\pi}   {\cal L}_{K\pi\pi} +
\alpha_{K_S \pi}   {\cal L}_{K_S \pi} 
\end{eqnarray}
The background fractions $\alpha_i$ depend on $\sqrt{s}$ and are taken
from Table \ref{tab:back_m3pi}.

\subsection{Results}
\label{sss-subresults}

In this section, we report on the fits to the substructure in 
$\tau^-\to \nu_\tau \pi^-\pi^0\pi^0$ decays.  
Given the complexity of the fitting procedure, we use 
only the lepton-tagged sample since the backgrounds from 
multihadronic events and other $\tau$ decays are smaller, 
particularly in the high $3\pi$ mass region.
We have performed many fits, including various amplitudes and 
employing differing assumptions.  Here, we present results from 
one fit based on the model described in Section~\ref{ss-modsub}, 
with certain parameters fixed as described below. 
Results obtained when 
these parameters were varied are given in Section~\ref{sss-submodvar}.

The resonances $Y$ shown in Table~\ref{tab:res_par} are implemented in the fit 
in amplitudes for $A \to Y \pi$, where $A$ represents an axial vector   
system.  
As mentioned above, we compute the normalization of the likelihood function 
in bins of $\sqrt{s}$ so as to be insensitive to the 
resonant content of the $3\pi$ system.  
In addition, the couplings $\beta_i$ could vary as a function of $\sqrt{s}$.  
This could be the case if, for example, 
several resonances contribute to the $3\pi$ system.  In our nominal 
fit, we constrain the $\beta_i$ to be independent of $\sqrt{s}$. 
For simplicity, we also consider the $1^+$ system 
to be point-like, {\sl i.e.}, we set $R_i = 0$ in Eqn.~\ref{eq:mesonradius},
with the result that $F_{R_i}=1$. 
Finally, we fix the $\nu_\tau$ helicity $h_{\nu_\tau}$ to its Standard 
Model value of $-1$.  Thus, our fit parameters consist of twelve real numbers: 
the moduli $\vert \beta_i \vert$ and phases $\phi_{\beta_i}$ 
of the couplings, for $i=2$--7.  Fits 
with $h_{\nu_\tau}$ floating are discussed in Section~\ref{ss-nutau}.  

\begin{figure}[th]
  \centering
  \leavevmode
  \epsfysize=16.cm
  \epsfbox{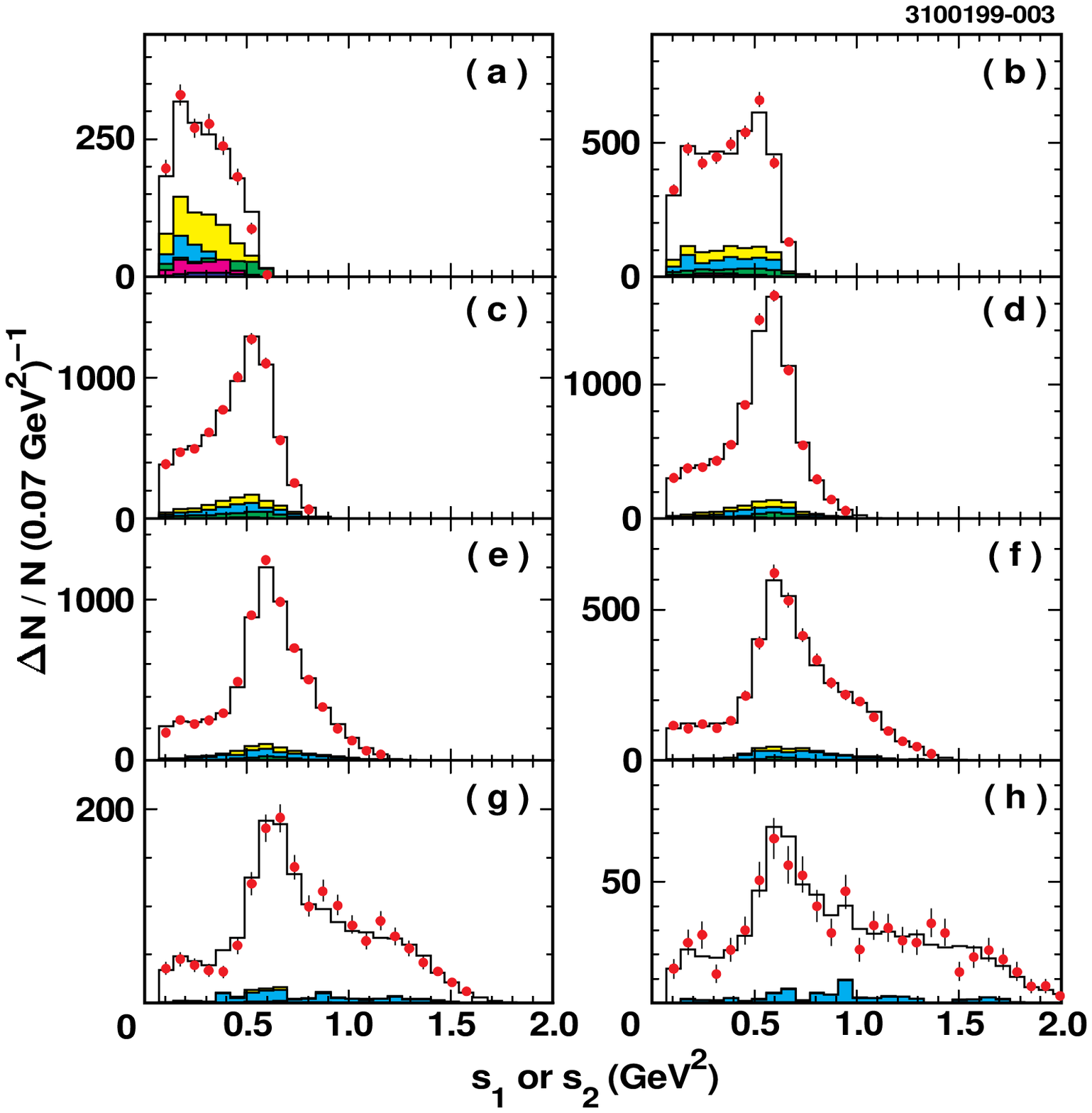}
  \caption[]{Dalitz plot projections:  
  distributions in squared $\pi^-\pi^0$ mass $s_1$ and $s_2$ 
  (two entries per event).
  The data are represented by the filled points. The solid
  line is the fit result. Background is represented by the shaded histograms.
  The lightest shaded histogram is the sum of the backgrounds, while the
  darker histograms show the backgrounds separately.
  Plots (a) to (h) correspond to slices in 
  $\sqrt{s}=$ 
  0.6-0.9, 0.9-1.0, 1.0-1.1, 1.1-1.2, 1.2-1.3, 1.3-1.4, 1.4-1.5, 1.5-1.8 
  $\mbox{GeV}$.}
  \label{s1s2_fitb}
\end{figure}

\begin{figure}[th]
  \centering
  \leavevmode
  \epsfysize=16.cm
  \epsfbox{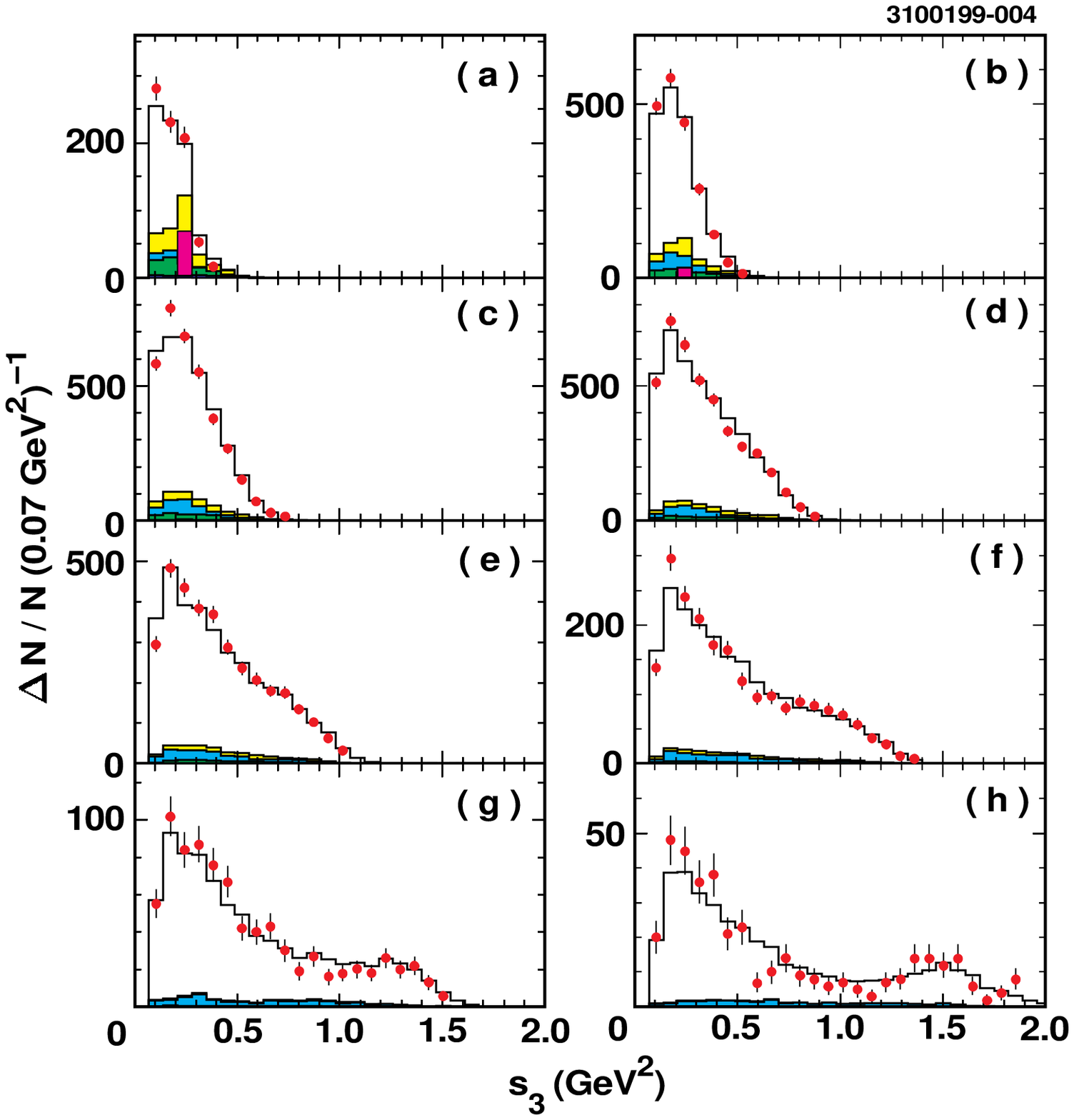}
  \caption[]{Dalitz plot projections:  
distributions in squared $\pi^0\pi^0$ mass $s_3$ (one entry per event).
The data are represented by the filled points. The solid
line is the fit result. Background is represented by the shaded histograms.
The lightest shaded histogram is the sum of the backgrounds, while the
darker histograms show the backgrounds separately.
Plots (a) to (h) correspond to slices in $\sqrt{s}=$ 
0.6-0.9, 0.9-1.0, 1.0-1.1, 1.1-1.2, 1.2-1.3, 1.3-1.4, 1.4-1.5, 1.5-1.8 
$\mbox{GeV}$.}
  \label{s3_fitb}
\end{figure}

\begin{figure}[th]
  \centering
  \leavevmode
  \epsfysize=16.0cm
  \epsfbox{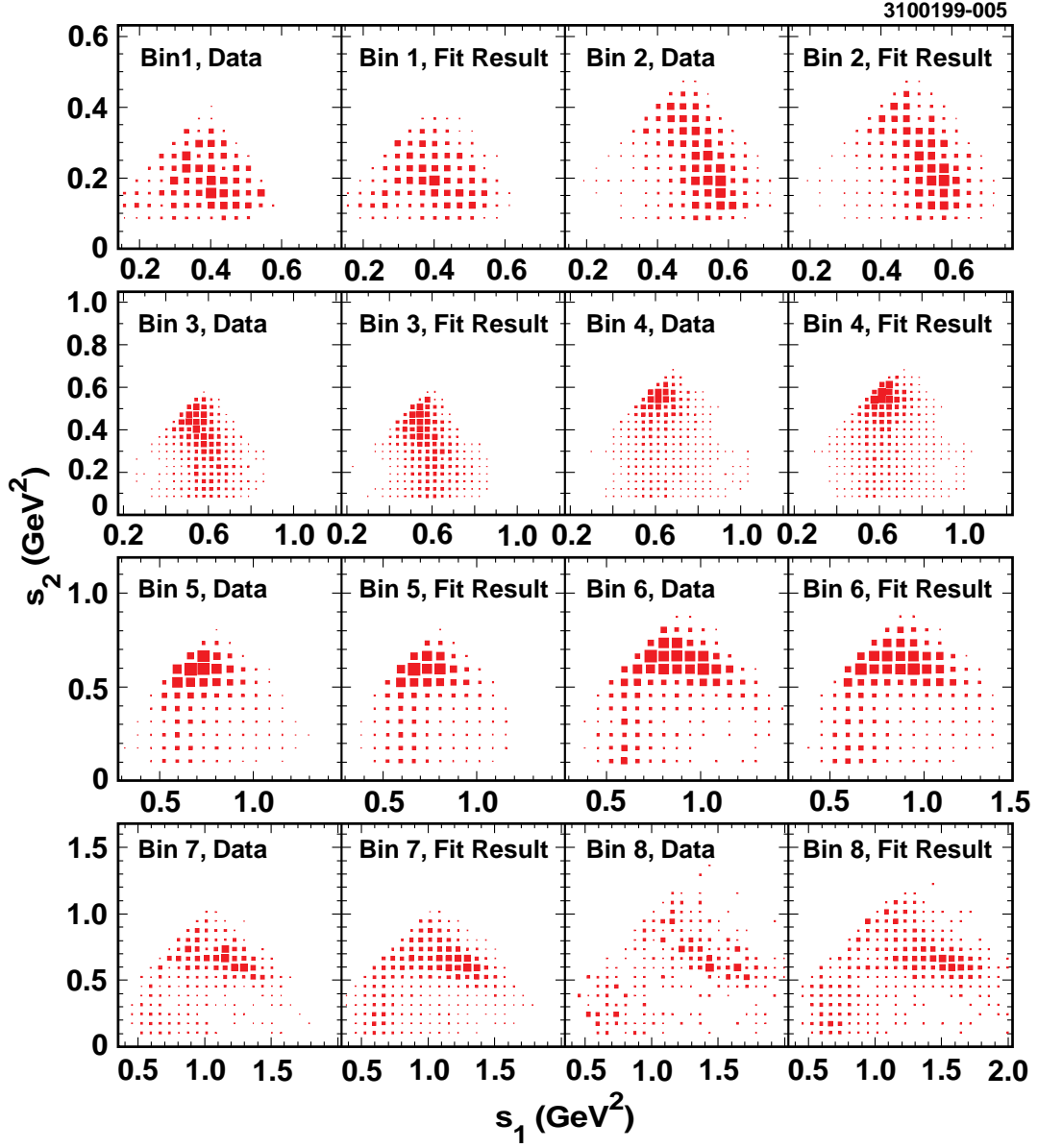}
  \caption[]{
Dalitz plot distributions for data and fit result.  Here $s_1$ is taken to 
be the higher of the two values of $M_{\pi^-\pi^0}^2$ in each event. 
Bins 1 through 8 correspond to slices in 
$\sqrt{s}=$ 
0.6-0.9, 0.9-1.0, 1.0-1.1, 1.1-1.2, 1.2-1.3, 1.3-1.4, 1.4-1.5, 1.5-1.8 
$\mbox{GeV}$.}
  \label{dal_fitb}
\end{figure}

The results from this nominal fit are summarized in 
Table~\ref{tab:result_r00}. 
The measured likelihood is 224259, while that expected is
$224340 \pm 214$.  The difference,  
$-0.4\,\sigma$, indicates an acceptable goodness of fit.  
As a function of $\sqrt{s}$, it is (in units of standard deviations): 
$+2.2$, $-1.3$, $-0.3$, $-1.5$, $+0.9$, $-0.7$, $-0.6$ and $+1.2$ in 
the eight slices of $\sqrt{s}$ defined in Table~\ref{tab:back_m3pi}.   
The significance of each amplitude  
is determined by repeating the fit with that amplitude excluded. 
Dalitz plot projections from the fit 
are shown in Figs.~\ref{s1s2_fitb} 
and~\ref{s3_fitb} in slices of $\sqrt{s}$, overlaid with the 
corresponding data distributions.  
The Dalitz plots themselves are shown in Fig.~\ref{dal_fitb}.  
A discussion of the results follows 
in Section~\ref{sss-subdisc}.  For now, we note the large 
contributions from channels involving isoscalars, in particular 
$\sigma\pi$ with a significance of $8.2 \sigma$.  

\begin{table}[hpbt]
\begin{center}
\caption[]{\small Results of the nominal fit for the moduli 
$\vert \beta_i\vert$ and phases $\phi_{\beta_i}$ of the coefficients 
for the amplitudes listed in Eqn.~\ref{eq:amplist}.
The two errors shown are statistical and systematic respectively.  
The branching fractions 
$\cal B$ are derived from the squared amplitudes  
(using the values of $\vert \beta_i\vert$), and are 
normalized to the total $\tau^-\to \nu_\tau \pi^-\pi^0\pi^0$ rate.  
These do not sum to 100$\%$, due 
to interference between the amplitudes.}
\label{tab:result_r00}        
\tabcolsep 5pt
{\small
\begin{tabular}{l l c  c c c}
& & Signif. 
  & $\vert\beta_i \vert$ & $\phi_{\beta_i}/\pi$
  & $\mbox{${\cal B}$ fraction}$$(\%)$
\\ \hline 
$\rho      $ & $s$-wave & ---  
             &$1$ & $0$ & $ 68.11 $ \\
$\rho(1450)$ & $s$-wave & $1.4\sigma$  
             &$0.12\pm 0.09\pm 0.03$ & $\phantom{-}0.99\pm 0.25\pm 0.04$
                                     & $ 0.30\pm 0.64\pm 0.17$  \\
$\rho      $ & $d$-wave & $5.0\sigma$  
             &$0.37\pm 0.09\pm 0.03$ & $          -0.15\pm 0.10\pm 0.03$ 
                                     & $ 0.36\pm 0.17\pm 0.06$ \\
$\rho(1450)$ & $d$-wave & $3.1\sigma$  
             &$0.87\pm 0.29\pm 0.06$ & $\phantom{-}0.53\pm 0.16\pm 0.06$ 
                                     & $ 0.43\pm 0.28\pm 0.06$ \\ 
$f_2 (1270)$ & $p$-wave & $4.2\sigma$  
             &$0.71\pm 0.16\pm 0.05$ & $\phantom{-}0.56\pm 0.10\pm 0.03$ 
                                     & $ 0.14\pm 0.06\pm 0.02$ \\ 
$\sigma    $ & $p$-wave & $8.2\sigma$  
             &$2.10\pm 0.27\pm 0.09$ & $\phantom{-}0.23\pm 0.03\pm 0.02$ 
                                     & $16.18\pm 3.85\pm 1.28$ \\ 
$f_0(1370)$  & $p$-wave & $5.4\sigma$  
             &$0.77\pm 0.14\pm 0.05$ & $          -0.54\pm 0.06\pm 0.02$ 
                                     & $ 4.29\pm 2.29\pm 0.73$ \\ 
\end{tabular}
}
\end{center}
\end{table}

\clearpage
\subsection{Modifications to the default model}
\label{sss-submodvar}

With the model described in Section~\ref{ss-modsub}, we have obtained a 
good fit to the Dalitz plot distributions.  
In this section we describe fits to variations of the model.  

\subsubsection{Uniformity of amplitude coefficients across $\sqrt{s}$}
\label{sss-subbetavar}

\begin{figure}[th]
  \centering
  \leavevmode
  \epsfysize=10.cm
  \epsfbox{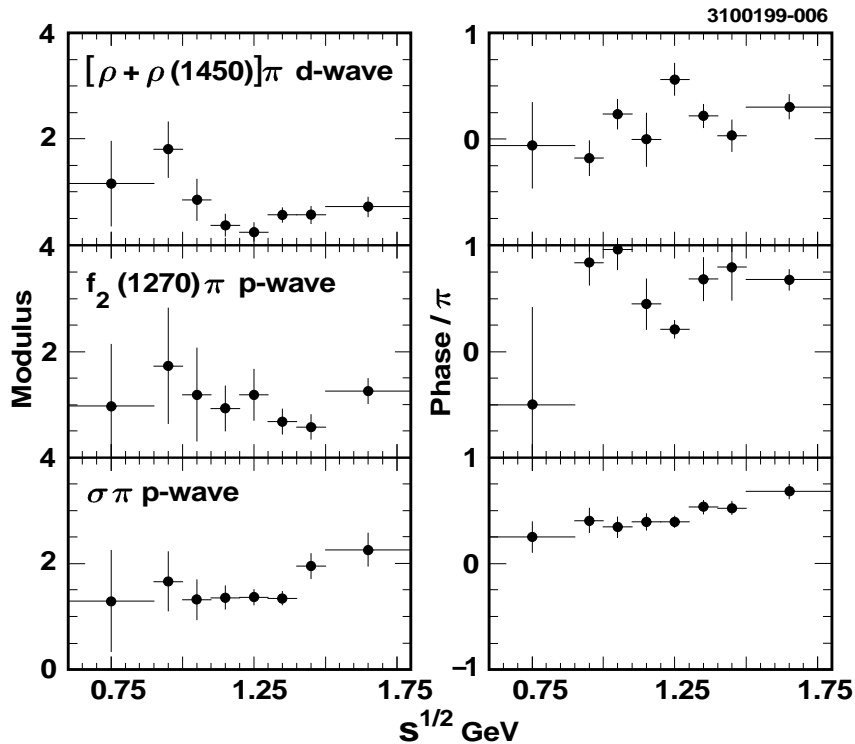}
  \caption[]{Results for the moduli and phases of coefficients 
           $\beta_i$ of amplitudes
           from a fit in which these are allowed to vary as a function of 
           the $3\pi$ invariant mass.  In this fit, only three amplitudes 
           are considered in addition to the $s$-wave $\rho\pi$ amplitude:
           $\rho\pi$ ($d$-wave), $f_2\pi$ ($p$-wave) and $\sigma\pi$ 
           ($p$-wave).}
  \label{fig:fita}
\end{figure}

The assumption that the coefficients $\beta_i$ for the 
various substructure amplitudes are independent of $\sqrt{s}$ may not 
be correct.  They would not be constant if, for example, more 
than one $3\pi$ resonance were present.  We have performed fits allowing 
the $\beta_i$ to vary; the results from one such fit are plotted in 
Fig.~\ref{fig:fita}.
In this fit, we considered fewer amplitudes so as to limit the number 
of fit parameters.  They are $\rho\pi$ ($d$-wave), $f_2\pi$ ($p$-wave) and 
$\sigma\pi$ ($p$-wave), in addition to the dominant $s$-wave $\rho\pi$ 
contribution.  Also, for this fit, we take the $\rho$ resonance to be 
the sum of $\rho(770)$ and $\rho(1450)$ amplitudes, with the $\rho(1450)$ 
admixture fixed according to studies of $\tau^-\to \nu_\tau \pi^-\pi^0$ 
decay~\cite{ju}.  The goodness of fit is acceptable: 
the measured likelihood minus that expected is $+0.2\sigma$.

The behavior of the moduli $|\beta_i|$ are consistent with uniformity 
across $\sqrt{s}$.  However, we note that the signficance of the $f_2\pi$ 
contribution is greatest in the highest mass slice.  We also see elevated 
contributions from the $\sigma\pi$ and $d$-wave $\rho\pi$ amplitudes in 
the high-mass slices, although these are not statistically significant.

\subsubsection{Importance of isoscalar contributions}
\label{sss-subisovar}

This analysis is the first study of the $a_1$ in $\tau$ decay 
to consider contributions from scalar $I=0$ mesons [$\sigma$ 
and $f_0(1370)$].  In addition, our fits return a significant 
$f_2(1270)\pi$ component.  Although these channels are expected 
to be present, a demonstration of the validity of the fit results 
is desirable given the complexity of both the model and the 
fit procedure.  To help visualize their 
collective importance in describing the substructure, 
we have performed fits excluding the three amplitudes $j_5$, 
$j_6$, and $j_7$ that involve isoscalars.  The Dalitz plot 
projections from one fit performed in this way are 
presented in Appendix~\ref{a-noiso} for comparison with those from 
the nominal fit.   We comment further on the impact of the 
large isoscalar contributions in Section~\ref{sss-subdisc}.

\subsubsection{Consideration of finite size of the $a_1$ meson}
\label{sss-subRvar}

\begin{figure}[th]
  \centering
  \leavevmode
  \epsfysize=8.0cm
  \epsfbox{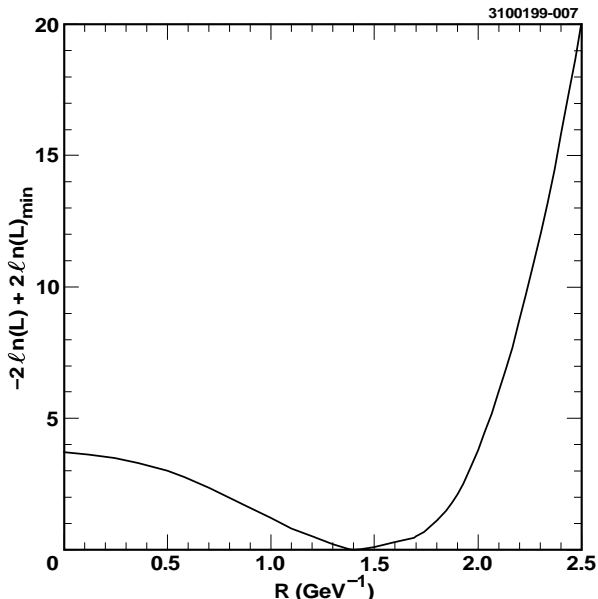}
  \caption[]{Dependence of $-2\ln {\cal L}$ on the 
  value of the meson radius parameter $R_i = R$ in variations of the nominal 
  ($R_i=0$) substructure fit.}
  \label{fig:Rvar}
\end{figure}

In our nominal fit, we set the $a_1$ radius to zero, such that the form 
factors $F_{R_i}$ in Eqn.~\ref{eq:mesonradius} are just unity.  
We find that good fits can also be obtained 
with non-zero values for $R_i$.  In Fig.~\ref{fig:Rvar}, we plot the 
differences in the minus log likelihood values from fits in which 
all $R_i$ are set to some value $R$.  
The best fit is obtained with $R = 1.4\,{\rm GeV}^{-1}$.  
We present the results from this fit in Table~\ref{tab:subfitr14} in 
Appendix~\ref{a-mesonradius}.
We will return 
to this issue in the context of the fits to the $3\pi$ mass spectrum 
in Section~\ref{ss-massfit}.

\subsubsection{Inclusion of $0^- \to 3\pi$ amplitudes}
\label{sss-subpscalars}

We have also performed fits including pseudoscalar contributions, namely 
$\pi^\prime \to \rho\pi$ ($p$-wave) and $\pi^\prime \to \sigma\pi$ ($s$-wave).
These amplitudes will necessarily have a different $\sqrt{s}$-dependence 
from those associated with axial vector production.  To account for this, 
we assume a 
Breit-Wigner form for the $\pi^\prime$ with a constant mass and width of 
1.300 and 0.400 GeV respectively, and use the results from one of the 
fits presented in Section~\ref{ss-massfit} for the $a_1$ parameters.  

In fits with each of these amplitudes included separately we find no  
statistically significant contributions.  We obtain the following 
90$\%$ CL limits:
\begin{eqnarray}
  {\cal B} (\tau\to\nu\pi^\prime \to \nu\rho\pi \to \nu 3 \pi) &
    < & 1.0\times 10^{-4},  \\
  {\cal B} (\tau\to\nu\pi^\prime \to \nu\sigma\pi \to \nu 3 \pi) &
    < & 1.9\times 10^{-4}.  
\end{eqnarray}

\subsubsection{Inclusion of other $1^+ \to 3\pi$ amplitudes}
\label{sss-subother}

In addition to the 
axial vector amplitudes $j^\mu_1,\ldots,j^\mu_7$, we performed 
fits including a contribution from $f_0(980)\pi$ ($p$-wave).  None 
of these fits returned a signficant coupling for this amplitude.  
Further discussion of possible $f_0(980)$ contributions 
appears in Section~\ref{sss-massmodother}.

\subsubsection{Variation of $\sigma$ meson resonance parameters}
\label{sss-subsigma}

By virtue of its low mass and large width, 
there is much uncertainty regarding the resonant shape 
of the $\sigma$ meson.  For simplicity,  
we have elected to characterize it using a Breit-Wigner form, with its
mass and width taken from the model of T\"ornqvist~\cite{tornscal}.  
We have not considered alternative forms, or explored extensively 
the range of possible resonance parameters.  However, in view of the large 
$\sigma$ contribution observed in this analysis, we have attempted to 
ascertain whether our data are sensitive to variation of its properties. 

We have refitted the data with a range of input values for the mass 
and width of the $\sigma$.  Of the values we considered, the best fit 
was obtained with $m_{0\sigma} = 555$~MeV and 
$\Gamma_{0\sigma} = 540$~MeV.  The value of $-2\ln {\cal L}$ for this fit 
was 224216.   This is 43 units below that for the nominal fit, but is 
still consistent with expectations given the statistics of the data 
sample.  
Using the smaller values of $m_{0\sigma}$ and 
$\Gamma_{0\sigma}$ has an impact on the values of $\beta_i$ obtained.
The main trend is a relative change of 20--40$\%$ 
in the $a_1$ branching fractions, which are smaller 
for the $\sigma\pi$ and $f_0(1370)\pi$ channels and are larger for 
the $\rho^\prime\pi$ and $f_2(1270)\pi$ channels.

\subsection{Systematic errors}

The systematic errors shown in Table~\ref{tab:result_r00} are based 
on estimates of the uncertainties arising from the following sources: 
Monte Carlo statistics, background fractions and modeling, dependence 
of the acceptance on the kinematical observables used in the fit, 
and detector resolution.  The uncertainties due to these sources are 
given in Table~\ref{tab:sys_error_r00}.
\begin{table}
\begin{center}
\caption[]{\small Systematic errors on hadronic substructure for the 
nominal fit.}
\label{tab:sys_error_r00}        
\tabcolsep 5pt
{\small
\begin{tabular}{c l  c c c c c c}
& & 
$\rho(1450)$& $\rho$     & $\rho(1450)$& $f_2 (1270) $& $\sigma$& $f_0 (1370)$\\
& & 
s-wave      & d-wave     & d-wave      & p-wave       & p-wave     & p-wave      \\
\hline
& $\Delta( \mbox{${\cal B}$ fraction})$ $(\%)$ &
$\pm 0.162$ & $\pm 0.058$& $\pm 0.045$ & $\pm 0.017$  & $\pm 1.201$& $\pm 0.701 $\\
\raisebox{1.4ex}[-1.4ex]{Monte Carlo}  & $\Delta( \vert \beta_i \vert)        $ & 
$\pm 0.029$ & $\pm 0.028$& $\pm 0.044$ & $\pm 0.047$  & $\pm 0.081$& $\pm 0.049$ \\
\raisebox{1.4ex}[-1.4ex]{statistics}   & 
$\Delta( \phi_{\beta_i} / \pi)$ &
$\pm 0.038$ & $\pm 0.027$& $\pm 0.058$ & $\pm 0.026$  & $\pm 0.011$& $\pm 0.020$ \\
\hline
& $\Delta( \mbox{${\cal B}$ fraction})$ $(\%)$ &
$\pm 0.027$ & $\pm 0.008$& $\pm 0.042$ & $\pm 0.014$  & $\pm 0.391$& $\pm 0.157$ \\
background                             & $\Delta( \vert \beta_i \vert)        $ & 
$\pm 0.006$ & $\pm 0.005$& $\pm 0.037$ & $\pm 0.010$  & $\pm 0.036$& $\pm 0.018$ \\
                                       & 
$\Delta( \phi_{\beta_i} / \pi)$ &
$\pm 0.005$ & $\pm 0.007$& $\pm 0.018$ & $\pm 0.009$  & $\pm 0.007$& $\pm 0.006$ \\
\hline
& $\Delta( \mbox{${\cal B}$ fraction})$ $(\%)$ &
$\pm 0.024$ & $\pm 0.014$& $\pm 0.006$ & $\pm 0.002$  & $\pm 0.178$& $\pm 0.097$ \\
efficiency                             & $\Delta( \vert \beta_i \vert)        $ & 
$\pm 0.004$ & $\pm 0.007$& $\pm 0.008$ & $\pm 0.002$  & $\pm 0.016$& $\pm 0.007$ \\
                                       & 
$\Delta( \phi_{\beta_i} / \pi)$ &
$\pm 0.013$ & $\pm 0.002$& $\pm 0.009$ & $\pm 0.001$  & $\pm 0.006$& $\pm 0.008$ \\
\hline
& $\Delta( \mbox{${\cal B}$ fraction})$ $(\%)$ &
$\pm 0.018$  & $\pm 0.006$ & $\pm 0.004$  & $\pm 0.002$   & $\pm 0.083$ & $\pm 0.065$  \\
\raisebox{1.4ex}[-1.4ex]{detector}     & $\Delta( \vert \beta_i \vert)        $ & 
$\pm 0.001$  & $\pm 0.002$ & $\pm 0.007$  & $\pm 0.001$   & $\pm 0.011$ & $\pm 0.006$  \\
\raisebox{1.4ex}[-1.4ex]{resolution}   & 
$\Delta( \phi_{\beta_i} / \pi)$ &
$\pm 0.002$  & $\pm 0.004$ & $\pm 0.006$  & $\pm 0.002$   & $\pm 0.004$ & $\pm 0.005$  \\
\hline
& $\Delta( \mbox{${\cal B}$ fraction})$ $(\%)$ &
$\pm 0.17$  & $\pm 0.06$ & $\pm 0.06$  & $\pm 0.02$   & $\pm 1.28$ & $\pm 0.73$  \\
total                                  & $\Delta( \vert \beta_i \vert)        $ & 
$\pm 0.03$  & $\pm 0.03$ & $\pm 0.06$  & $\pm 0.05$   & $\pm 0.09$ & $\pm 0.05$  \\
                                       & 
$\Delta( \phi_{\beta_i} / \pi)$ &
$\pm 0.04$  & $\pm 0.03$ & $\pm 0.06$  & $\pm 0.03$   & $\pm 0.02$ & $\pm 0.02$  \\ 
\end{tabular}
}
\end{center}
\end{table}

The error due to Monte Carlo statistics is based on the variance of results 
obtained from six separate fits, each using one sixth of the 
Monte Carlo sample for the normalization of the likelihood function.  
Fits performed after 
varying the background fractions and model (in the case of the 
$\nu_\tau \,\pi^-\,3\pi^0$ channel) within reasonable limits were used 
to estimate the error associated with this source.  To estimate the 
error associated with acceptance, the Monte Carlo was used to parameterize 
acceptance as a function of the charged and neutral pion momenta as well 
as the opening angles between these particles.  Reasonable deviations 
from these parameterizations were used to reweight events entering the 
fit, and the resulting variations in fit parameters were taken as the 
systematic errors.  Finally, the likelihood function in Eqn.~\ref{eq:likel} 
does not take into account resolution effects.  The effects of modifying 
it to include resolution smearing based on errors in track parameters 
for the $\pi^-$ and in photon energies and directions for the $\pi^0$'s  
was used to estimate the error from this source.  For all fit parameters, 
the error due to limited Monte Carlo statistics dominates the systematic 
error.  

The results given in Table~\ref{tab:result_r00} are meaningful 
only in the context of the model used to parametrize the substructure.   
Different models yield results that can differ significantly from our 
nominal fit results.  Given this plus the unfeasibility of examining all 
possible models, we have not attempted to assign a systematic error 
associated with model dependence.

\subsection{Discussion}
\label{sss-subdisc}

The results of the fits for the substructure can be summarized 
as follows:
\begin{itemize}
\item The $\rho\pi$ $s$-wave amplitude with a branching fraction of around
      $70\%$ is dominant, as expected.
\item With the exception of the $\rho(1450)\pi$ $s$-wave 
      amplitude, all amplitudes included in the nominal fit 
      contribute significantly to the $3\pi$ hadronic current. In other 
      fits, we find no evidence for contributions from $a_1\to f_0(980)\pi$, 
      or from $\tau^-\to\nu_\tau \pi^{\prime -}$.  
\item The isoscalar mesons $f_2$, $f_0(1370)$, and $\sigma$ contribute
      with a combined branching fraction of approximately $20\%$ to the 
      $3\pi$ hadronic width. In particular, the $\sigma$ meson with
      a significance of $\sim 8\sigma$ cannot be neglected.  It 
      shows up strongly as part of the broad enhancement at the low 
      end in the $s_3$ ($M_{\pi^0\pi^0}^2$) distribution in the slices 
      of $\sqrt{s}$ shown in Fig.~\ref{s3_fitb}(g) and (h), 
      but is significant in all slices.
\item The $\rho(1450)\pi$ state shows up more strongly in the $d$-wave 
      amplitude than in the $s$-wave amplitude.        
\end{itemize}

The last point above may have implications regarding a possible $a_1^\prime$ 
contribution.  One expects that an $a_1^\prime$ contribution induces
$\sqrt{s}$-dependent couplings.  
In the fits allowing $\beta_i$ to vary with $\sqrt{s}$, we 
found that (1) the goodness of fit is not significantly better, and 
(2) the values of $\beta_i(s)$ are roughly consistent with being constant.  
On the other hand, according to the flux-tube-breaking model of 
Refs.~\cite{KI,GI}, 
the $a_1^\prime$ meson prefers to decay to $\rho\pi$ by $D$-wave rather 
than $S$-wave, and the $\rho(1450)$ is preferred over the
$\rho(770)$.  Thus, it is possible that the 
measured $\rho(1450)\pi$ $d$-wave amplitude could be 
induced by an $a_1^\prime$.
The suggestions of enhanced $f_2\pi$ and $\sigma\pi$ contributions 
at large $\sqrt{s}$ are also consistent with the 
hypothesis of an $a_1^\prime$.
However, the statistics of the present data sample 
are not sufficient to resolve this question with the substructure fits. 

\begin{figure}[th]
  \centering
  \leavevmode
  \epsfysize=14.0cm
  \epsfbox{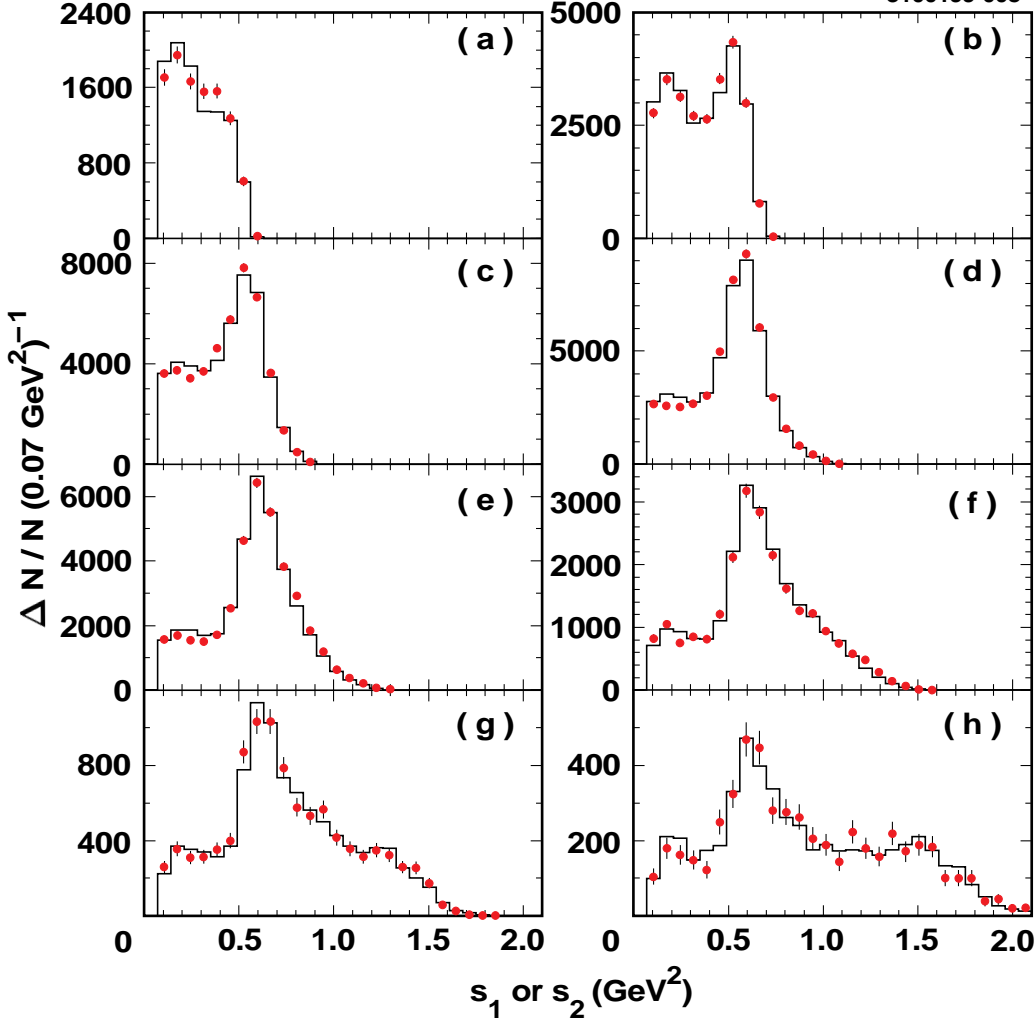}
  \caption[]{Background-subtracted squared $\pi^+\pi^-$ mass 
  ($s_1$ and $s_2$) spectrum for the 
  three charged pion mode (two entries per event).
  The data are represented by the filled points. The solid
  line is the isospin prediction based on 
  our fit to the $\pi^-\pi^0\pi^0$ mode.}
  \label{s1s2_charged}
\end{figure}

As a test of our fit results, we have compared the Dalitz 
plot distributions from a sample of background-subtracted 
$\tau^-\rightarrow\nu_\tau \pi^-\pi^-\pi^+$ events 
with the isospin prediction based on the results from the 
nominal fit to the $\pi^-\pi^0\pi^0$ mode.  The backgrounds 
were estimated from the generic $\tau$ Monte Carlo sample.  The 
dominant backgound $\tau^-\to\nu_\tau \pi^-\pi^-\pi^+\pi^0$ 
is simulated in this sample with the model implemented 
in {\tt TAUOLA}, containing
$\pi^-\omega$ as well as $[\rho\pi\pi]^-$ (in various charge 
combinations) substructure.  
The Dalitz plot projections are shown in Fig.~\ref{s1s2_charged}.   
The observation that the hadronic current for the all-charged mode 
is well described by our results for the $\pi^-\pi^0\pi^0$ mode 
provides a critical corroboration of our measurements.  This is  
particularly important for the amplitudes involving isoscalars 
since they enter the all-charged mode with the opposite sign 
relative to the other amplitudes.  A full analysis of 
the high-statistics all-charged mode is under way and will be 
presented in the future.  

The world average values~\cite{pdg98} for the 
$\tau^-\to\nu_\tau \pi^-\pi^+\pi^-$ and 
$\tau^-\to\nu_\tau \pi^-\pi^0\pi^0$ branching fractions 
are $(9.23\pm 0.11)\,\%$ and $(9.15\pm 0.15)\,\%$, respectively.  
Their near equality is consistent with expectations 
from isospin symmetry, if the decays were to proceed 
exclusively via $\rho\pi$ or $\rho^\prime\pi$.  
One consequence of the presence of isoscalars in $a_1$ decay 
is the possibility of upsetting this expectation.
However, due to interference 
the two modes contribute nearly equally to $\Gamma^{a_1}_{tot}(s)$ 
(see Fig.~\ref{fig:all_m3pi_n}(b) in Section~\ref{ss-massresults}).  
The divergence of these contributions at high values 
of $\sqrt{s}$ is damped in the decay rate 
by the falling of the $a_1$ line shape, as well as by 
phase space and weak interaction dynamics in the $\tau$ decay.  
Furthermore, 
the residual preference for the all-charged mode at high $\sqrt{s}$ is 
compensated for by the larger phase space available for the $\pi^-2\pi^0$ 
mode at low $\sqrt{s}$.  Quantitatively, 
the ratio of branching fractions is predicted from this analysis
to be ${\cal B}(\tau^-\to\nu_\tau \pi^-\pi^+\pi^-)/
       {\cal B}(\tau^-\to\nu_\tau \pi^-\pi^0\pi^0) = 0.985$, 
in agreement with the ratio $1.009 \pm 0.020$ obtained from the direct 
measurements.  

Finally, the branching fractions reported in Table~\ref{tab:result_r00}
are the $\tau^-$ decay branching fractions relative to the total 
$\nu_\tau \pi^-\pi^0\pi^0$ rate.  These differ from the 
$a_1$ branching fractions due to the weighting of the $a_1$ line shape 
by factors associated with $\tau^-\to \nu_\tau a_1^-$ weak decay.  
The $a_1$ branching ratios 
  ${\cal B}(a_1^-\to [Y\pi]^- \to \pi^-\pi^0\pi^0) /
   {\cal B}(a_1^-\to \pi^-\pi^0\pi^0)$ 
are given in Table~\ref{tab:result_a1br}. 
\begin{table}
\begin{center}
\caption[]{\small Branching ratios for $a_1^-$ decay into $\pi^-\pi^0\pi^0$
via intermediate states shown, relative to the total 
$a_1^-\to \pi^-\pi^0\pi^0$ rate.  The errors shown are statistical only.}
\label{tab:result_a1br}        
\begin{tabular}{ccc}
\multicolumn{2}{c}{Amplitude}  & Branching ratio $(\%)$ \\
\hline 
$\rho      \pi$ & $s$-wave & $                   60.19$ \\
$\rho(1450)\pi$ & $s$-wave & $ \phantom{1}0.56\pm 0.84$ \\
$\rho      \pi$ & $d$-wave & $ \phantom{1}1.30\pm 0.60$ \\
$\rho(1450)\pi$ & $d$-wave & $ \phantom{1}2.04\pm 1.20$ \\
$f_2 (1270)\pi$ & $p$-wave & $ \phantom{1}1.19\pm 0.49$ \\
$\sigma    \pi$ & $p$-wave & $           18.76\pm 4.29$ \\
$f_0(1370) \pi$ & $p$-wave & $ \phantom{1}7.40\pm 2.71$ \\
\end{tabular}
\end{center}
\end{table}

\section{DETERMINATION OF THE SIGNED \boldmath $\nu_\tau$ HELICITY}
\label{ss-nutau}

In the fits reported in the previous section, the $\tau$ neutrino 
(anti-neutrino) helicity was fixed to the Standard Model value of 
$h_{\nu_\tau} = -1$ ($h_{\overline{\nu}_\tau} = +1$).  
However  
as first pointed out by K\"uhn and Wagner~\cite{KW}, interference 
between the two $\pi^-\pi^0$ systems gives rise to a parity violating 
term in the squared matrix element for the decay 
$\tau^-\to\nu_\tau \pi^-\pi^0\pi^0$. 
This permits determination of the sign, as well as the magnitude, of the 
neutrino helicity. 
Including $h_{\nu_\tau}$ as an additional free parameter to the nominal 
fit described in the previous section, 
and assuming invariance under the 
combined charge-conjugation and parity ($CP$) operation, 
we find $ h_{\nu_\tau} = -1.00 \pm 0.12 $.   In this fit, 
the values of $\beta_i$ are affected at a negligible level.  
Including only the $s$-wave $\rho\pi$ amplitude in the model for 
the substructure yields a poor fit, with $h_{\nu_\tau} = -0.73\pm 0.09$.  

To investigate the model dependence entering this measurement 
we also performed fits for substructure and $h_{\nu_\tau}$ with different 
input parameters.  While our nominal fit was performed with the $a_1$ 
radius set to zero, non-zero values also gave good fits, as noted in 
the previous section.  Using the best fit value of 
$R_i = 1.4 \mbox{ GeV}^{-1}$, we obtain
$  h_{\nu_\tau} = -1.03 \pm 0.13 \, .$ 
As a best estimate of $h_{\nu_\tau}$ given the dependence on 
input assumptions, we average this value with the $R_i = 0$ result to 
obtain 
\begin{equation}
  h_{\nu_\tau} = -1.02 \pm 0.13 \, \mbox{\sl (stat.)} 
                       \pm 0.01 \, \mbox{\sl (syst.)} 
                       \pm 0.03 \, \mbox{\sl (model)}\, ,  
\end{equation}
where the uncertainty due to the model dependence is estimated by the 
difference between the values from these two fits.  This result agrees 
with other determinations~\cite{pdg98} of the sign and 
magnitude of $h_{\nu_\tau}$, 
as well as with the Standard Model value of $-1$.  

The systematic error given for $h_{\nu_\tau}$ was determined in a fashion 
similar to those in the substructure analysis.  The sources contributing 
to this error are: Monte Carlo statistics ($\pm 0.005$), background 
determination ($\pm 0.010$), dependence of the acceptance on the kinematic 
observables used in the fit ($\pm 0.003$), and detector resolution 
($\pm 0.004$).

We have also looked for possible $CP$ non-conservation by 
determining $h_{\nu_\tau}$ and $h_{\overline{\nu}_\tau}$ 
separately.  Defining a $CP$-violating asymmetry 
\begin{equation}
  A_{CP} = \frac{  h_{\overline{\nu}_\tau}  +  h_{\nu_\tau}  }
                { |h_{\overline{\nu}_\tau}| + |h_{\nu_\tau}| }, 
\end{equation}
we find $A_{CP} = -0.08\pm 0.13$, where the error is dominantly due to 
statistics.

\section{ANALYSIS OF THE \boldmath $\pi^-\pi^0\pi^0$ MASS SPECTRUM}
\label{ss-massfit}

In this section, we describe fits to the $3\pi$ mass spectrum 
performed to extract resonance parameters of the $a_1$ meson.  
The results from the substructure fits are used as inputs for 
determination of the running of the $a_1$ mass and width.

\subsection{Fitting method and assumptions}
\label{ss-massfitmethod}

The $a_1$ resonance parameters are determined from a $\chi^2$ fit  
to the background-subtracted and efficiency-corrected $\pi^-\pi^0\pi^0$ 
mass spectrum.  The fit is performed over a range from 
$M_{3\pi} = 0.600$ to 1.725 GeV in bins of width 25 MeV.
As in the substructure analysis, we perform a variety 
of fits, reflecting different models and assumptions.  We perform fits 
to the all-tagged sample, as well as to the lepton-tagged subsample, to 
benefit from the higher statistics.  

For our nominal fit, we specify the following version of the model
described in Section~\ref{ss-modmass}:
(1) turn off  $a_1^\prime$ contribution, {\sl i.e.}, set $\kappa = 0$
        in Eqn.~\ref{eq:breit};
(2) turn off form factors describing finite size of the $a_1$, 
        {\sl i.e.}, set $R_i=0$; 
(3) include the $K^*K$ threshold, but not the $f_0(980)\pi$ threshold,  
        in determining $\Gamma^{a_1}_{tot}(s)$; and 
(4) assume the running $a_1$ mass to be flat as a function of $\sqrt{s}$.
We have also performed fits in which various of these specifications 
are modified, and 
obtain satisfactory results under a variety of configurations. 
With the above specifications, the fit contains three free parameters in 
addition to an overall normalization: the pole $a_1$ mass $m_{0a_1}$,
the $a_1\to 3\pi$ coupling $g_{a_1(3\pi)}$, 
and the relative $a_1\to K^*K$ coupling $\gamma_{a_1(K^*K)}$.  We 
derive from these parameters 
the pole $a_1$ width $\Gamma_{0a_1}$ using Eqn.~\ref{eq:runwidth}.

Some comments on the above choices are in order.  The 
$\sqrt{s}$-dependence of the $a_1$ total width depends 
strongly on assumptions.  
Inclusion of the $K^*K$ channel is motivated by observation of the decay 
$\tau^-\to \nu_\tau [K^*K]^-$~\cite{pdg98}, although it is not 
well-determined as to how much of this 
comes through the axial vector (rather than vector) weak current.  
As we have no evidence for the $f_0(980)\pi$ channel in the substructure 
fits, we have omitted its possible contribution in our nominal fit here.  
However, this and other thresholds may be present.  For 
example, a possible $a_1\to f_1(1285)\pi$ channel, as suggested by the 
recent obervation of this system in $\tau$ decay~\cite{vasilii},  
would open up near the $K^*K$ threshold.  Thus, the value 
for $\Gamma^{a_1}_{K^*K}$ returned from our fit can not be strictly 
interpreted as just the $a_1\to K^*K$ partial width.  

The running of the $a_1$ mass is even 
more problematic since the upper limit of integration (over $ds^\prime$) 
in Eqn.~\ref{eq:disper} is infinity, 
and thus the integral will include effects from channels that open above 
the $\tau$ mass, and are therefore not directly measureable in $\tau$ 
decay.  Furthermore, a damping of the amplitudes, such as that provided 
by the form factors $F_{R_i}$, is needed so that the integral can converge.  
As a result, 
allowing the $a_1$ mass to run is practical only in models where 
the $R_i$ are non-zero.  In such models, the effect of additional thresholds 
at high $\sqrt{s}$ is to flatten the $\sqrt{s}$-dependence of the 
running mass.  Thus, we expect the running mass to be closer to a constant 
than we would predict from Eqn.~\ref{eq:runmass} with known thresholds.  
Setting $R_i=0$ and taking a constant $a_1$ mass may not 
be rigorous, however the resulting model is simplified.  

\subsection{Results}
\label{ss-massresults}

The results obtained from our nominal fit to the $3\pi$  resonance 
shape parameters are shown in the second column of Table~\ref{tab:result_m3pi}.
The $\chi^2$ for this fit is 39.3 for 41 degrees of freedom.  
Fits to just the lepton-tagged event sample yield consistent results.

\begin{figure}[hbpt]
  \centering
  \leavevmode
  \epsfysize=6.5in
  \epsfbox{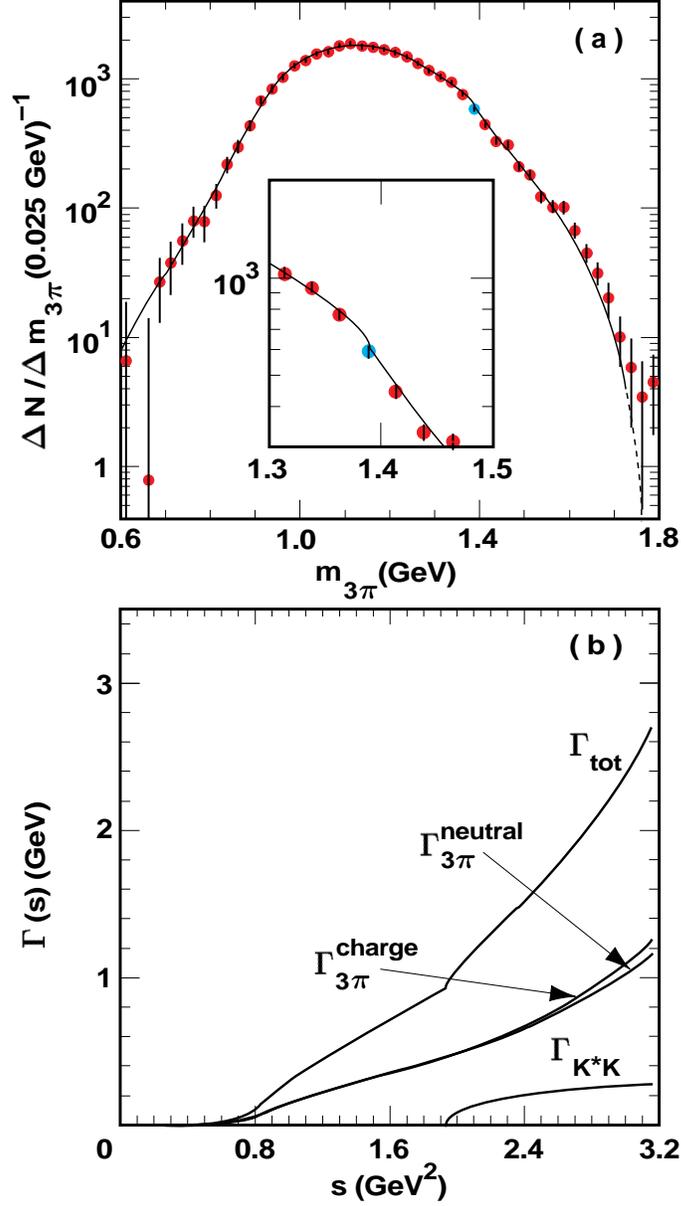}
  \caption[]{(a) Background-subtracted, efficiency-corrected $3\pi$ mass
spectrum from the all-tag sample. The solid line shows the result 
from the nominal fit.  The inset shows a magnified view of the 
$a_1 \to K^*K$ threshold region.  
In (b), the $\sqrt{s}$-dependent 
$a_1$ width $\Gamma_{tot}^{a_1}(s)$ is plotted as a function of $s$, 
along with the separate contributions from the channels considered, 
as given in Eqn.~\ref{eq:runwidth}.  The labels 
$\Gamma^{\rm charge}_{3\pi}$ and $\Gamma^{\rm neutral}_{3\pi}$ refer 
to the contributions from the $a_1^-\to \pi^-\pi^+\pi^-$ and 
$\pi^-\pi^0\pi^0$ channels respectively.  
}
  \label{fig:all_m3pi_n}
\end{figure}

The background-subtracted, efficiency-corrected $3\pi$ mass spectrum 
from the all-tagged sample is shown in Fig.~\ref{fig:all_m3pi_n}(a), 
with the function corresponding to the nominal fit overlaid.
Shown in Fig.~\ref{fig:all_m3pi_n}(b) is the $a_1$ width 
$\Gamma_{tot}^{a_1}(s)$, as defined by Eqn.~\ref{eq:runwidth}, as well 
as the contributions from the individual $a_1$ decay channels considered.
The kink associated with the turn-on of $a_1\to K^*K$ is 
visible in the $3\pi$ mass spectrum at $\sim 1.375$~GeV.  
\begin{table}[htbp]
\begin{center}
\caption[]{Results for fits to the $3\pi$ resonance shape,  
with nominal background and efficiency correction.  The second column 
gives results based on the nominal fit function.  The first error is 
that due to statistics, while the second is the systematic error.  The 
third column gives results from the fit including an $a_1^\prime$ 
contribution, with statistical errors only.  The derived quantity 
$ {\cal B} (K^* K )$, the $a_1\to K^*K$ branching fraction, is also shown
assuming this is the only amplitude accounted for by 
the threshold function labeled $\Gamma_{K^*K}$ in 
Fig.~\ref{fig:all_m3pi_n}(b) (see text).  
The quantity $\phi_\kappa$ denotes the phase of the $a_1^\prime$ 
amplitude relative to that of the $a_1$.}
\label{tab:result_m3pi}
\begin{tabular}{l c c}
 Fit Parameter  & Nominal Fit & Fit with $a_1^\prime$ \\
\hline
$ m_{0a_1}                    \mbox{ (GeV)}$       & $1.331\pm 0.010\pm 0.003$ 
                                                   & $1.330\pm 0.011$ \\
$\Gamma_{0a_1}                  \mbox{ (GeV)}    $ & $0.814\pm 0.036\pm 0.013$ 
                                                   & $0.814\pm 0.038$ \\
$\gamma_{a_1 (K^\star K)}                        $ & $3.32 \pm 0.26 \pm 0.04 $ 
                                                   & $3.72 \pm 0.45$ \\
$ {\cal B} (K^\star K )$ $(\%)                   $ & $3.3  \pm 0.5  \pm 0.1  $ 
                                                   & $4.0  \pm 1.0$ \\
$ |\kappa | $                                    & 0   & $0.053 \pm 0.019$ \\
$ \phi_\kappa/\pi $                              & --- & $0.10 \pm 0.22$ \\
$\chi^2/ndof $                                     & $39.3/41$ & $28.9/39$ \\
\end{tabular}
\end{center}
\end{table}
%

\subsection{Modifications to the nominal fit function}

\subsubsection{Finite size of the $a_1$ meson}

Like the Dalitz plot distributions, the $3\pi$ mass spectrum contains some 
sensitivity to the parametrization of the $a_1$ as being point-like 
or of finite size.  Here the sensitivity depends also on the 
treatment of the $\sqrt{s}$-dependence of $a_1$ mass.  

As in the substructure analysis, we re-fit the data multiple times, 
stepping through a range of values for the $a_1$ meson size parameter $R$.    
The results from these fits are given in Appendix~\ref{a-mesonradius}.
For both the all-tag and lepton-only-tag samples, the best fits were obtained 
with values of $R$ between 1.2 and 1.4~GeV$^{-1}$, depending on whether 
the $a_1$ mass was treated as a running or constant mass.  
This agrees well with the substructure analysis 
which favors a value of $R$ of 1.4~GeV$^{-1}$.  
This value of $R$ corresponds to an r.m.s.~radius of the $a_1$ of 0.7~fm.  
It is interesting to note that this is similar to the value 
employed by Isgur, Morningstar and Reader~cite{IMR} in their analysis of 
the $3\pi$ line shapes from the DELCO~\cite{delco}, 
MARK II~\cite{markii} and ARGUS~\cite{arg86} experiments.
However, the statistics 
of the present sample are not sufficient to determine the 
necessity of including the $F_R(k)$ form factor in the parametrization 
of the hadronic current.  

Despite this, uncertainty on this issue represents a 
substantial source of model dependence with regard to the $a_1$ resonance 
parameters.  Tables~\ref{tab:mass_rvar_c} and \ref{tab:mass_rvar_r} in 
Appendix~\ref{a-mesonradius} demonstrate this point.  As an example, 
for $R=1.2\,\mbox{GeV}^{-1}$ with a constant $a_1$ mass, we find 
$m_{0a_1} = 1.285\pm 0.007$~GeV and $\Gamma_{0a_1} = 0.619\pm 0.021$~GeV   
(statistical errors only).

\subsubsection{Running of the $a_1$ mass}

As indicated above, 
we have also performed fits with the $\sqrt{s}$-dependence of the 
$a_1$ mass computed according to Eqn.~\ref{eq:runmass}.  This can 
only be done in models with non-zero values for the $a_1$ size 
parameter $R$.  
The results, also given in Appendix~\ref{a-mesonradius}, indicate that 
slightly better fits can be obtained using a running $a_1$ mass.  However, 
since satisfactory fits are obtained with a constant mass, we 
conclude that the present data sample is not sensitive to the 
running of the $a_1$ mass.

\subsubsection{Inclusion of an $a_1^\prime(1700)$ admixture}

\begin{figure}[thb]
  \centering
  \leavevmode
  \epsfysize=6.5in
  \epsfbox{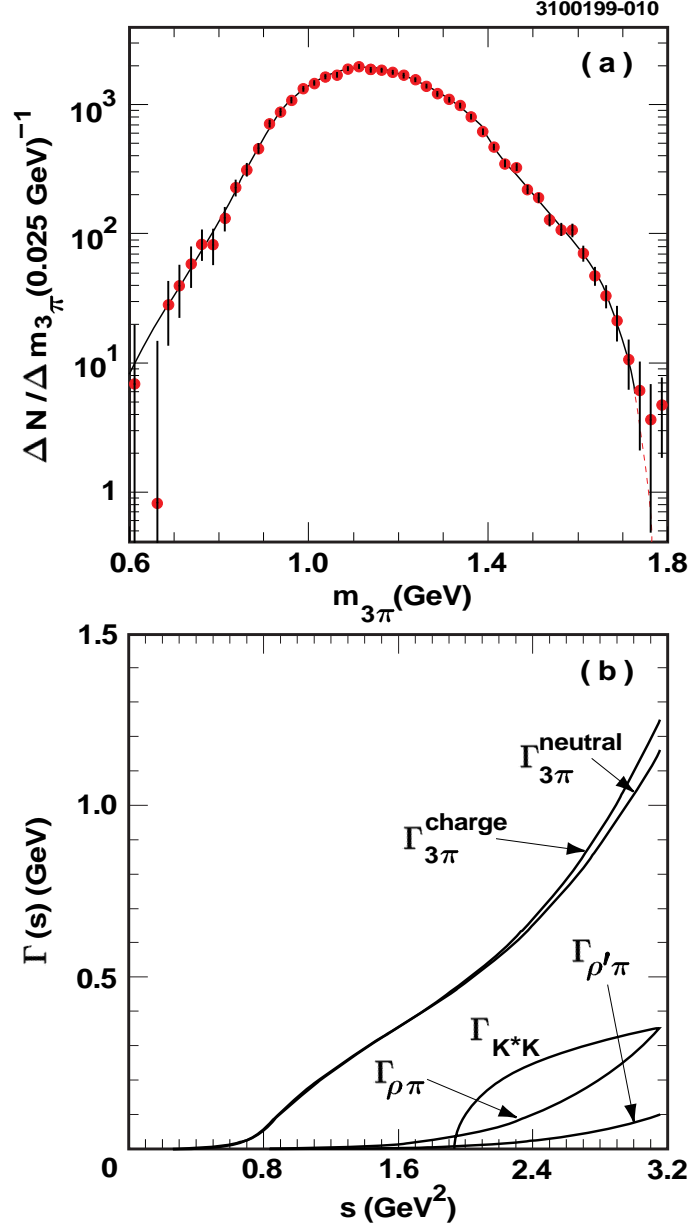}
  \caption[]{(a) Background-subtracted, efficiency-corrected $3\pi$ mass
spectrum from the all-tag sample. The solid line shows the 
fit result for the model in  
which an $a_1^\prime$ admixture is allowed to contribute.
In (b) the assumed $\rho\pi$ and $\rho^\prime\pi$ contributions 
to the $a_1^\prime$ width in this model are shown, 
along with the contributions to the $a_1$ width from 
Fig.~\ref{fig:all_m3pi_n} for comparison.
}
  \label{fig:all_m3pi_a1p}
\end{figure}

Despite the goodness of the fit to the nominal model, 
the data above 1.575 GeV show an excess 
relative to the fit function in Fig.~\ref{fig:all_m3pi_n}.  This region 
is where contributions from interference with an $a_1^\prime$ meson with 
mass around 1.7~GeV might appear.  We have performed various 
fits allowing $\kappa$ in Eqn.~\ref{eq:breit} to float.
The results from one such fit are given in the last column of 
Table~\ref{tab:result_m3pi} and plotted in Fig.~\ref{fig:all_m3pi_a1p}(a).  
In this fit, we 
have used $m_{0a_1^\prime} = 1.700$~GeV and $\Gamma_{0a_1^\prime} = 
0.300$~GeV.  We have also fixed the $a_1^\prime(\rho^\prime\pi)$ coupling 
to be equal to the $a_1^\prime(\rho\pi)$ coupling to determine 
$\Gamma_{a_1^\prime}(s)$ as shown in Fig.~\ref{fig:all_m3pi_a1p}(b).   
This is an {\sl ad hoc} choice, however the fit is relatively 
insensitive to the parametrization of $\Gamma_{a_1^\prime}(s)$.  

The $\chi^2$ for this fit is 28.9 for 39 degrees of freedom. The 
$a_1^\prime$ contribution has a significance of 2.8$\sigma$, 
with $\vert \kappa\vert = 0.053\pm 0.019$ and a phase $\phi_\kappa$ 
consistent with zero.  If we take $\phi_\kappa$ to be zero, 
the resulting fit yields $\vert\kappa\vert = 0.060\pm 0.017$.  
To test whether backgrounds in the all-tag sample are 
influencing this result, we have also fit the spectrum 
from just the lepton-tag events (with $\phi_\kappa$ floating). 
This fit yields consistent 
results, with $|\kappa | = 0.05\pm 0.04$.  

The analysis of DELPHI~\cite{delphi} yielded a large amplitude for 
the $a_1^\prime$ contribution, with $|\kappa|$ in the range of $0.50\pm 0.06$ 
to $0.75\pm 0.06$, depending on the model used.  The corresponding values 
obtained here are smaller by an order of magnitude, and are much less 
significant statistically.  However, the models used by DELPHI include 
neither the contributions to $\Gamma^{a_1}_{tot}(s)$ from the opening of 
the $K^*K$ channel, nor those associated with the isoscalar channels. 
In particular, the $K^*K$ amplitude has a significant effect at large 
values of $\sqrt{s}$ where the spectrum is most sensitive to the presence 
of an $a_1^\prime$.  Omitting the $K^*K$ amplitude in our fit, we obtain a 
value of $\vert\kappa\vert$ approximately twice as large as our nominal 
value.  Finally, the DELPHI anaylsis involves a simultaneous fit to 
the $3\pi$ mass spectrum and the Dalitz plot projections at large 
$\sqrt{s}$, employing assumptions for the substructure 
composition that differ from what we have determined (for the $a_1$) in 
our substructure analysis.  
In summary, direct comparison of our value of 
$\kappa$ with that from the DELPHI analysis is not meaningful.

\subsubsection{Opening of $f_0(980)\pi$ threshold}
\label{sss-massmodother}

Since the impact of the opening of $K^*K$ threshold appeared to be 
significant, we also performed fits including the opening of $f_0(980)\pi$ 
threshold.  This would include contributions through both $f_0\to \pi\pi$ 
and $K\overline{K}$ channels.  The $f_0\to \pi^0\pi^0$ channel would 
be expected to contribute to the substructure in our $\pi^-\pi^0\pi^0$ 
sample, but the other modes would not do so.

Marginal improvements in the fit quality were obtained only 
for those models in which the $a_1$ mass was run according to 
Eqn.~\ref{eq:runmass}.  In these cases the 
$f_0(980)\pi$ contributions to the total width of the $a_1$ were 
present at the $2\sigma$ level or less.  Fits including this contribution 
are discussed in Appendix~\ref{a-mesonradius}. 

To summarize, we find no evidence for contribution from the opening of 
the $a_1\to f_0(980)\pi$ channel in either the substructure fits or 
the $3\pi$ mass fits.  However, other scalar $I=0$ mesons [$\sigma$ and 
$f_0(1370)$] are needed to provide a good description of the substructure.  
This observation may have some bearing on the interpretation 
of the $f_0(980)$ as something other than a $q\overline{q}$ meson, 
as has been frequently speculated (see for example Ref.~\cite{pdg98}).  
Given the theoretical and experimental complexity, we 
cannot comment on this issue except to 
note that the non-observation of the $f_0(980)$ in $a_1$ decay 
is not inconsistent with an exotic interpretation for this state.

\subsection{Systematic errors}
\label{ss-masssys}

The sources of the systematic errors shown in Table~\ref{tab:result_m3pi} 
are just those associated with background subtraction and acceptance.
The background errors are estimated by repeating the fit after 
separately varying the amount of each of the backgrounds being subtracted. 
We vary the background fractions of modes with two real $\pi^0$'s 
by three times the uncertainty on their branching fractions. 
In the case of the fake-$\pi^0$ background, the subtraction is varied by 
three times the statistical error of the $\pi^0\pi^0$ side band sample. 
Finally, the $\sqrt{s}$-dependence of the acceptance as determined 
from Monte Carlo events is parametrized; these parameters are then 
varied by three standard deviations to estimate the associated uncertainty.

The systematic errors shown in Table~\ref{tab:result_m3pi} are dominated 
by the errors due to the background subtraction.  
As a check, we have performed fits allowing the 
separate background normalization and 
acceptance correction functions to float, subject to constraints added 
to the $\chi^2$ on the magnitude of their deviations from nominal.  The 
changes in fit parameters observed in these fits are small relative to the 
quoted errors.  The deviations of the correction functions from nominal 
are also small in this fit.  As in the substructure fits, we do not 
assign a systematic error for model dependence.

\subsection{Discussion}

As can be seen from Fig.~\ref{fig:all_m3pi_n}(a), and from the 
$\chi^2$ for the nominal fit shown, 
the model described in Sec.~\ref{ss-modmass},  
with the assumptions listed in Sec.~\ref{ss-massfitmethod}, 
provides a good description of the data.  
The $a_1$ pole mass and width are determined precisely, however 
their values depend significantly on the model and input assumptions. 

The results demonstrate 
the importance of including the opening of $K^*K$ threshold.  
Fits performed without this contribution to $\Gamma^{a_1}_{tot}(s)$ 
yield large $\chi^2$ values.  
The obtained $a_1\to K^*K$ branching fraction $(3.3 \pm 0.5) \%$ 
corresponds to a $\tau$ branching fraction of 
${\cal B}(\tau^-\to \nu_\tau a_1^- \to \nu_\tau [K^*K]^-) 
= (0.16 \pm 0.03)\%$.  
For comparison,  
multiplying the directly measured $\tau^-\to\nu_\tau K^{*0}K^-$ 
branching fraction~\cite{pdg98}    
by a factor of two (to account for $K^{*-}K^0$) gives 
${\cal B}(\tau^-\to \nu_\tau [K^*K]^-) = (0.42\pm 0.08)\%$.
The apparent shortfall in our measurement 
is not surprising since the $\nu_\tau [K^*K]^-$ final 
state is expected to receive contributions from both vector 
and axial vector hadronic currents. 

In the high-mass region above 1.575 GeV, the data appear to 
be systematically high relative to the fit function.  To understand 
whether this is an experimental effect, we have modified 
the background and efficiency corrections within reasonable limits.  
As described in the previous section, we 
have also performed fits in which background and efficiency 
correction functions are allowed to float.   
Neither of these approaches significantly 
improves the fit in this region.  
Fits with non-zero values of $R_i$, with 
and without a running $a_1$ mass, and/or including the $f_0(980)\pi$ 
threshold also fail in this respect.  

Including a contribution from an $a_1^\prime$ meson, however, visibly 
influences the shape of the $3\pi$ mass spectrum in the high-mass 
region, as shown in Fig.~\ref{fig:all_m3pi_a1p}(a).  
As noted earlier, the presence of an 
$a_1^\prime$ may also be consistent with an enhanced $d$-wave 
contribution from $\rho(1450)\pi$ relative to $s$-wave, as observed 
in the substructure analysis.  
We have not evaluated systematic errors associated with the $a_1^\prime$ 
contribution determined from the $3\pi$ mass spectrum fits, 
since these are likely dominated by uncertainties associated 
with the modeling of the $a_1$ line shape.  
We have performed other fits with an $a_1^\prime$, 
sampling the range of model variations described 
above.  The statistical significance of the $a_1^\prime$ contribution 
is typically 2-3$\,\sigma$ in these fits. 
We conclude that more data is needed to establish 
whether the $a_1^\prime$ is present.  
%

\section{SUMMARY AND CONCLUSIONS}
\label{s-conclude}

To summarize, we have presented a detailed 
model-dependent analysis of hadronic structure 
in the decay $\tau^- \to \nu_\tau \pi^-\pi^0\pi^0$ using data obtained 
with the CLEO~II detector.  This decay mode represents a unique source 
of information on the $I=1$ axial vector meson sector, an area 
of hadron spectroscopy which is difficult to 
access cleanly via other production mechanisms.  
In this analysis we have derived successful descriptions of both 
the $a_1$ line shape and the substructure present in its decay to 
three pions.  
The most significant result is the observation of large contributions 
to the substructure from intermediate states involving the 
isoscalar mesons $\sigma$, $f_0(1370)$, and $f_2(1270)$.  
With this, our data also provides new input on the complicated 
$I=0$ scalar meson sector:  we observe some sensitivity to the 
properties of the $\sigma$ meson, for example.  
More generally, significant progress towards a 
satisfactory description of Dalitz plot distributions in 
$\tau^-\to \nu_\tau [3\pi]^-$ decay has been achieved.  
This is supported 
by the observation that our characterization of substructure in the 
$\nu_\tau \pi^-\pi^0\pi^0$ mode also provides a good description of 
substructure in the $\nu_\tau \pi^-\pi^+\pi^-$ data.  

Using the results from the substructure fits 
to infer the $\sqrt{s}$-dependence of the $a_1$ width, 
we have determined the $a_1$ meson resonance parameters.  We obtain 
$m_{0a_1} = 1.331 \pm 0.010 \pm 0.003$ GeV and 
$\Gamma_{0a_1} = 0.814\pm 0.036\pm 0.013$ GeV, although these values 
depend significantly on the details of the model used to fit the $3\pi$ 
mass spectrum.  For example, 
taking the meson size parameter to be 
$R=1.2\,\mbox{GeV}^{-1}$ instead of zero, we find 
$m_{0a_1} = 1.285 \pm 0.007\,\mbox{(stat.)}$~GeV and 
$\Gamma_{0a_1} = 0.619\pm 0.021\,\mbox{(stat.)}$~GeV.  
Such model dependences are not reflected in the quoted 
systematic errors.  We also find a significant contribution to the 
$\sqrt{s}$-dependence of the $a_1$ width associated with the opening 
of the $a_1\to K^*K$ decay channel at high values of $\sqrt{s}$.  

We have investigated the possibility of an additional 
contribution to the $3\pi$ mass spectrum from a radially excited 
$a_1^\prime$ meson, as suggested by the analysis of 
DELPHI~\cite{ronan,delphi}.  This is also suggested by our data, 
which show an apparent 
excess of events at large $3\pi$ mass relative to various fits without 
an $a_1^\prime$ component.  The data are better described 
with an $a_1^\prime$ contribution, 
though at a level below that reported by DELPHI. 
The model used in our analysis differs substantially 
from that analysis, with regard to treatment of the $K^*K$ 
threshold and the substructure in the $a_1\to 3\pi$ channel.  
We have not assessed the impact on 
$\nu_\tau$ mass studies of effects associated with 
the complex substructure in $a_1\to 3\pi$ decay or 
the apparent distortions in the $3\pi$ mass spectrum caused by the 
$K^*K$ and possible $a_1^\prime$ contributions.  However,   
careful consideration of such effects in the course of these analyses 
should improve the reliability of the ensuing 
tau neutrino mass constraints.  

We have also obtained a precise determination of the signed 
$\tau$ neutrino helicity 
$h_{\nu_\tau} = -1.02 \pm 0.13\;(\mbox{stat.}) 
                      \pm 0.03\;(\mbox{syst.}+\mbox{model}) $ 
when this quantity is left as a free parameter in the substructure fits.  
As has been noted in earlier measurements of this quantity~\cite{arg93}, 
accurate parameterization of the substructure is important for obtaining 
an unbiased measurement.  With the improved understanding of this 
substructure, this result provides unambiguous evidence for the 
left-handedness of the $\tau$ neutrino.

We have addressed several additional issues pertaining to the 
characterization of axial vector meson decay dynamics.  
For example, the data show limited sensitivity 
to the finite size of the $a_1$ meson, both in the substructure and 
the $3\pi$ mass spectrum fits.  With the parametrization of the 
associated form factor used here, we find that both analyses favor 
an r.m.s.~$a_1$ radius of around 0.7 fm.  As with the question of a 
possible $a_1^\prime$ contribution, a definitive conclusion on this 
issue requires additional data.  We have looked for indications 
of non-axial-vector contributions to the $3\pi$ substructure via 
the $\pi^\prime(1300)$ resonance, and have placed upper limits on the 
$\tau$ decay rate to this state.  
Detailed analyses of the higher-statistics 
$\tau^- \to \nu_\tau \pi^-\pi^+\pi^-$ data should shed 
additional light on these and other issues. 

Although the quantitative results presented in this article are 
strongly model-dependent, they describe successfully the qualitative 
features of the data.  However, further insight can be gained from 
a quantitative model-independent analysis of the data, such as that 
proposed by K\"uhn and Mirkes~\cite{KM}.  We are presently pursuing 
such an analysis and will report the results in a separate article.

\section*{ACKNOWLEDGEMENTS}
We gratefully acknowledge the effort of the CESR staff in providing us with
excellent luminosity and running conditions.
J.R. Patterson and I.P.J. Shipsey thank the NYI program of the NSF, 
M. Selen thanks the PFF program of the NSF, 
M. Selen and H. Yamamoto thank the OJI program of DOE, 
J.R. Patterson, K. Honscheid, M. Selen and V. Sharma 
thank the A.P. Sloan Foundation, 
M. Selen and V. Sharma thank Research Corporation, 
S. von Dombrowski thanks the Swiss National Science Foundation, 
and H. Schwarthoff thanks the Alexander von Humboldt Stiftung for support.  
This work was supported by the National Science Foundation, the
U.S. Department of Energy, and the Natural Sciences and Engineering Research 
Council of Canada.

\clearpage 
\appendix

\section{PARAMETERIZATION OF SUBSTRUCTURE IN \boldmath 
         $\tau^-\to \nu_\tau \pi^- \pi^0 \pi^0$}
\label{a-substructure}

To parameterize the model used to fit for the substructure in 
$\tau^-\to\nu_\tau \pi^-\pi^0\pi^0$, 
we follow many of the conventions used in the {\tt KORALB/TAUOLA}~\cite{korb} 
$\tau\tau$ Monte Carlo generator.  
We first denote the four-momenta of the pions by
\begin{eqnarray}
p_1 &=&\mbox{ four-momentum of $\pi^0_1$,}\nonumber\\
p_2 &=&\mbox{ four-momentum of $\pi^0_2$, and}\nonumber\\
p_3 &=&\mbox{ four-momentum of $\pi^-$.} 
\end{eqnarray}
We then define the quantities 
$q_1 = p_2 - p_3$, $q_2 = p_3 - p_1$, $q_3 = p_1 - p_2$, and 
$a = p_1 + p_2 + p_3$.  Ignoring for now the resonant structure of the 
$3\pi$ system, the general form for $3\pi$ hadronic current $J^\mu$ as 
defined by Eqs.~\ref{eq:taudec} and~\ref{eq:current} can be written as:
\begin{equation}
\label{eq:koralorg}
J^\mu = T^{\mu\nu} 
\left[   c_1  q_{1\nu}  F_1 + c_2  q_{2\nu}  F_2 
       + c_3  q_{3\nu}  F_3 \right] 
       + c_4  a^\mu F_4  
       + c_5  i \epsilon^{\mu\nu\rho\sigma}p_{1\nu}p_{2\rho}p_{3\sigma}
              F_5 \, ,
\end{equation}
where the $c_i$ are complex scalar coefficients, and 
$T^{\mu\nu} =  g^{\mu\nu} - {a^\mu a^\nu}/{a^2} $.  

The form factors $F_i$ contain the description of the low energy QCD 
phenomena we are studying.  
Owing to Lorentz invariance, they depend only on Lorentz scalars.
The terms containing $F_1$, $F_2$ and $F_3$ are associated with the 
axial vector contribution.  The term containing $F_3$ is not needed since 
it can be absorbed into the terms containing $F_1$ and $F_2$.  
We write it explicitly here to make connection with form factors 
associated with specific resonant substructure. $F_4$ is the scalar 
form factor, and $F_5$ is the $G$-parity violating vector form factor.  
Neither of these are expected to contribute significantly in 
$\tau^-\to\nu_\tau [3\pi]^-$, hence we generally set these terms to zero, 
except where noted.

\subsection{Model for the form factors}

The form factors as defined in Eqn.~\ref{eq:koralorg} do not have a 
simple correspondance with those that can be associated with 
specific resonant contributions to the hadronic current.  Here we give 
the ansatz for the amplitudes $j^\mu_i$ (as defined in Eqn.~\ref{eq:jmu}) 
of the hadronic current in
the decay $\tau^-\rightarrow\nu_\tau\pi^-\pi^0\pi^0$ 
used in the substructure fits: 
\begin{equation}
\label{eq:2pi0pi}
\arraycolsep0.1mm
\begin{array}{rccrcc}
j^\mu_1 &=& 
T^{\mu\nu}  
 & \Bigg[ &
 q_{1\nu}  B^P_\rho (s_1) F_{R_1} (k_1) - 
 q_{2\nu}  B^P_\rho (s_2) F_{R_1} (k_2)    
& \Bigg] \\ \rule[-2.0ex]{0pt}{10.0ex}
j^\mu_2 &=& 
T^{\mu\nu}  
 & \Bigg[ &
 q_{1\nu} B^P_{\rho^\prime} (s_1) F_{R_2} (k_1)  - 
 q_{2\nu} B^P_{\rho^\prime} (s_2) F_{R_2} (k_2)
& \Bigg] \\ \rule[-2.0ex]{0pt}{10.0ex}
j^\mu_3 &=& 
T^{\mu\nu}  
 & \Bigg[ &
Q_{1\nu} \left(a q_1\right) B^P_\rho  (s_1)  F_{R_3} (k_1) -  
Q_{2\nu} \left(a q_2\right) B^P_\rho  (s_2)  F_{R_3} (k_2)
& \Bigg] \\ \rule[-2.0ex]{0pt}{10.0ex}
j^\mu_4 &=& 
T^{\mu\nu}  
 & \Bigg[ &
Q_{1\nu} \left(a q_1\right) B^P_{\rho^\prime}  (s_1) F_{R_4} (k_1)   -  
Q_{2\nu} \left(a q_2\right) B^P_{\rho^\prime}  (s_2) F_{R_4} (k_2)
& \Bigg] \\ \rule[-2.0ex]{0pt}{10.0ex}
j^\mu_5 &=& 
T^{\mu\nu}  
 & \Bigg[ &
\phantom{+}\left[
q_{3\nu}  \left( a q_3 \right) - 
{\DD\frac{1}{3}} \left[ a_\nu - h_{3\nu}
{\DD\frac{\left(h_3 a \right)}{s_3}}\right] \left(q_3 q_3 \right) \right] 
B^D_{f_2}  (s_3)   F_{R_5} (k_3)
& \Bigg] \\ \rule[-2.0ex]{0pt}{10.0ex}
j^\mu_6 &=& 
T^{\mu\nu}  
& \Bigg[ &
Q_{3\nu}  B^S_{\sigma} (s_3)   F_{R_6} (k_3)
& \Bigg] \\ \rule[-2.0ex]{0pt}{10.0ex}
j^\mu_7 &=& 
T^{\mu\nu}  
 & \Bigg[ &
Q_{3\nu}  B^S_{f_{0} } (s_3)   F_{R_7} (k_3)
& \Bigg] \, .
\end{array}
\end{equation}
The kinematic factors appearing in above are defined as follows:
\begin{eqnarray}
h_i &  = &  p_j + p_k \mbox{ with ($i\ne j \ne k \ne i$)}\, ,   \nonumber\\
Q_i &  = &  h_i - p_i \, ,
\end{eqnarray}
\begin{equation}                                      
s_i = h_i^2 \, , \quad s = a^2 \, , 
\end{equation}                      
and the decay momenta are given by     
\begin{eqnarray}
\label{eq:dec}
k_i &  = &  \frac{ \sqrt{ 
\left( s-{\left( \sqrt{s_i} + m_i \right)}^2 \right)
\left( s-{\left( \sqrt{s_i} - m_i \right)}^2 \right) }}{2\sqrt{s}} \nonumber\\ 
k_i^\prime &  = &  \frac{ \sqrt{ 
\left( s_i -{\left( m_j + m_k \right)}^2 \right)
\left( s_i -{\left( m_j - m_k \right)}^2 \right) }}{2\sqrt{s_i}} 
\mbox{ with ($i\ne j \ne k \ne i$)} \, .
\end{eqnarray}
For the complex couplings $\beta_i$, 
we specify $\beta_1 =1$, and thus we determine the
remaining couplings relative to the first amplitude ($\rho\pi$, $s$-wave). 
The Breit-Wigner functions for the intermediate states are
\begin{eqnarray}
\label{eq:bwdef}
B^L_Y (s_i)        & = &
\frac{m_{0Y}^2}{(m^2_{0Y} - s_i) - i m_{0Y} \Gamma^{Y, L}(s_i)}  \,\,  \nonumber\\ 
\Gamma^{Y, L}(s_i) & =  & 
\Gamma^Y_0  {\left( \frac{ k^\prime_i }{  k^\prime_0 } \right)}^{2L+1}
\frac{m_{0Y}}{\sqrt{s_i}} \, .
\end{eqnarray}
The parameters  $m_{0Y}$, $\Gamma^Y_0$, $k_i^\prime$ and  
$k_0^\prime$ are the nominal mass, nominal width,
the decay momentum, and the decay momentum at $s_i = m^2_{0Y}$, respectively. 
See Table \ref{tab:res_par} for a summary of the resonance parameters of the 
intermediate states as used in the substructure fits.
The ansatz used for the form factors $F_{R_j}$ is
\begin{equation}
\label{eq:form}
F_{R_j} (k_i )  = \exp ( -\frac{1}{2} R^2_j k^2_i ) \, .
\end{equation}  

\subsection{Connection with reduced form factors}

With the ansatz $J^\mu = \sum_{i=1}^7 \beta_i \, j_i^\mu$,  
the reduced form factors $F_i$ of Eqn.~\ref{eq:koralorg} are given by:
\begin{eqnarray}
F_1 & = &    \phantom{-\frac{1}{3}}\beta_1 B^P_\rho (s_1)  F_{R_1} (k_1)         
\nonumber\\ \rule[-2.0ex]{0pt}{6.0ex}
    &   &   + \phantom{\frac{1}{3}}\beta_2 B^P_{\rho^\prime} (s_1)   F_{R_2} (k_1)
\nonumber\\ \rule[-2.0ex]{0pt}{6.0ex}
    &   &   - \frac{1}{3}\beta_3 ((s_3-m^2_3)-(s_1-m^2_1)) 
                                  B^P_\rho (s_2)  F_{R_3} (k_2)                   
\nonumber\\ \rule[-2.0ex]{0pt}{6.0ex}
    &   &   - \frac{1}{3}\beta_4 ((s_3-m^2_3)-(s_1-m^2_1)) 
                                  B^P_{\rho^\prime} (s_2)   F_{R_4} (k_2)        
\nonumber\\ \rule[-2.0ex]{0pt}{6.0ex}
    &   &   + \frac{1}{3}\beta_5 
         \frac{(a^2-m^2_3+s_3)(2 m^2_1 + 2 m^2_2 - s_3)}{6s_3}
               B^D_{f_2} (s_3)  F_{R_5} (k_3)                                    
\nonumber\\ \rule[-2.0ex]{0pt}{6.0ex}
    &   &   + \frac{2}{3}\beta_6 B^S_\sigma (s_3)   F_{R_6} (k_3)                
\nonumber\\ \rule[-2.0ex]{0pt}{6.0ex}
    &   &   + \frac{2}{3}\beta_7 B^S_{f_0}  (s_3)   F_{R_7} (k_3)                
\\ \rule[-2.0ex]{0pt}{6.0ex}
F_3 & = & \phantom{+} \frac{1}{3}\beta_3 \left[ 
             ((s_2-m^2_2)-(s_3-m^2_3)) B^P_\rho (s_1) F_{R_3} (k_1)  + 
             ((s_3-m^2_3)-(s_1-m^2_1)) B^P_\rho (s_2) F_{R_3} (k_2)  
                               \right]                             
\nonumber\\ \rule[-2.0ex]{0pt}{6.0ex}
    &   &          +  \frac{1}{3}\beta_4 \left[ 
            ((s_2-m^2_2)-(s_3-m^2_3)) B^P_{\rho^\prime} (s_1) F_{R_4} (k_1) +  
            ((s_3-m^2_3)-(s_1-m^2_1)) B^P_{\rho^\prime} (s_2) F_{R_4} (k_2) 
                               \right]                             
\nonumber\\ \rule[-2.0ex]{0pt}{6.0ex}
    &   &          -  \frac{1}{2}\beta_5  
            ((s_1-m^2_1)-(s_2-m^2_2)) B^D_{f_2} (s_3) F_{R_5} (k_3) 
\\ \rule[-2.0ex]{0pt}{6.0ex}  
F_4 & = & 0 
\end{eqnarray}
$F_2$ is given by $F_1$ under an interchange of indices $1 \leftrightarrow 2$, 
with a relative minus sign between $F_1$ and $F_2$ required 
({\sl i.e.}, the coefficients $c_i$ Eqn.~\ref{eq:koralorg} 
follow $c_2 = - c_1$).


\section{SUBSTRUCTURE FITS EXCLUDING ISOSCALARS}
\label{a-noiso}

\begin{figure}[th]
  \centering
  \leavevmode
  \epsfysize=16.cm
  \epsfbox{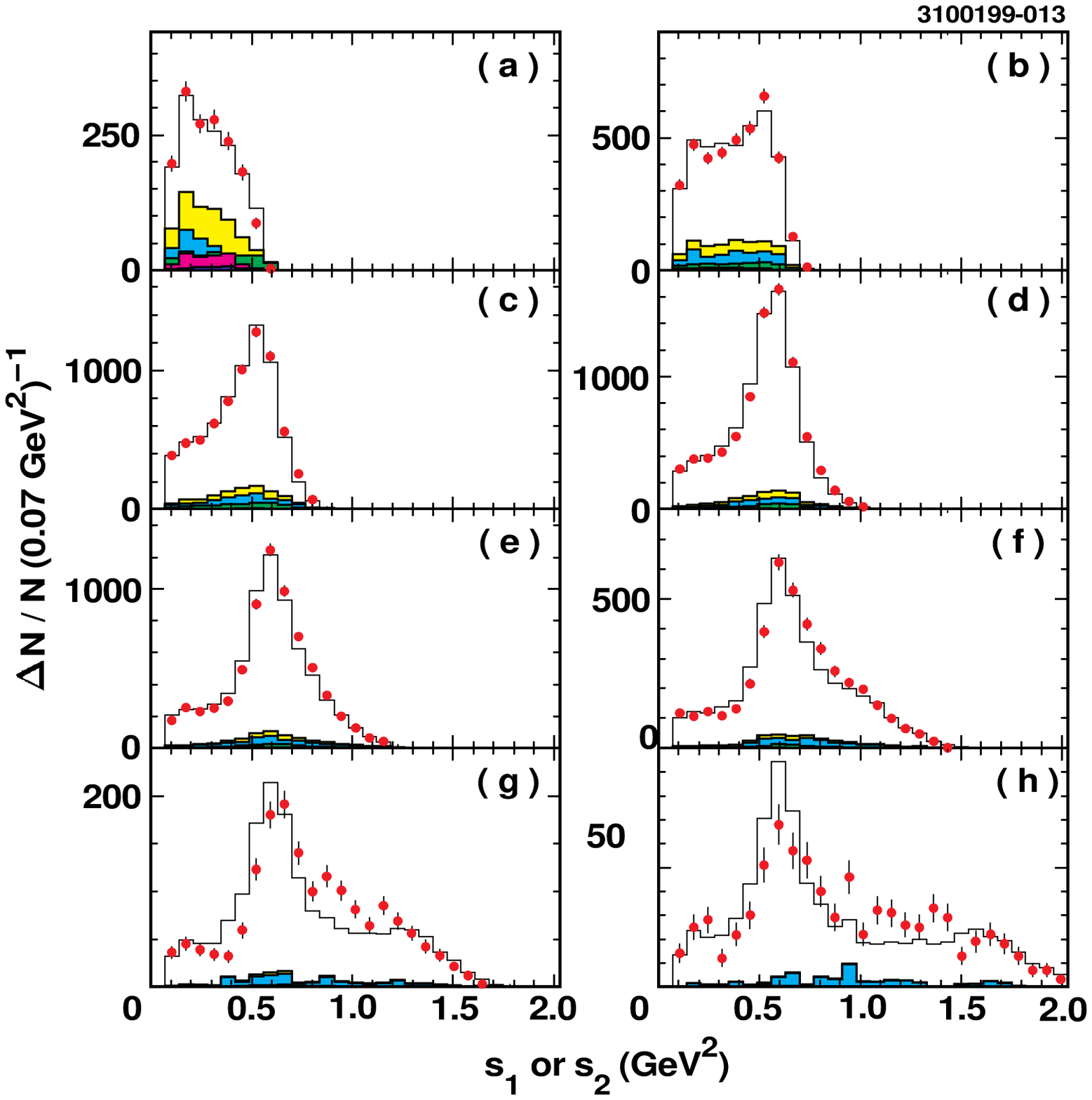}
\caption[]{Dalitz plot projections: 
distributions in squared $\pi^-\pi^0$ mass $s_1$ and $s_2$ 
(two entries per event), plotted for slices of $\sqrt{s}$  
as in Fig.~\ref{s1s2_fitb}.
The data are represented by the filled points. The solid
line is the result from a modified version of the nominal fit 
in which amplitudes involving isoscalar mesons 
($\sigma\pi$, $f_0\pi$ and $f_2\pi$) have been omitted.  
}
\label{fig:s1s2_noiso}
\end{figure}

\begin{figure}[th]
  \centering
  \leavevmode
  \epsfysize=16.cm
  \epsfbox{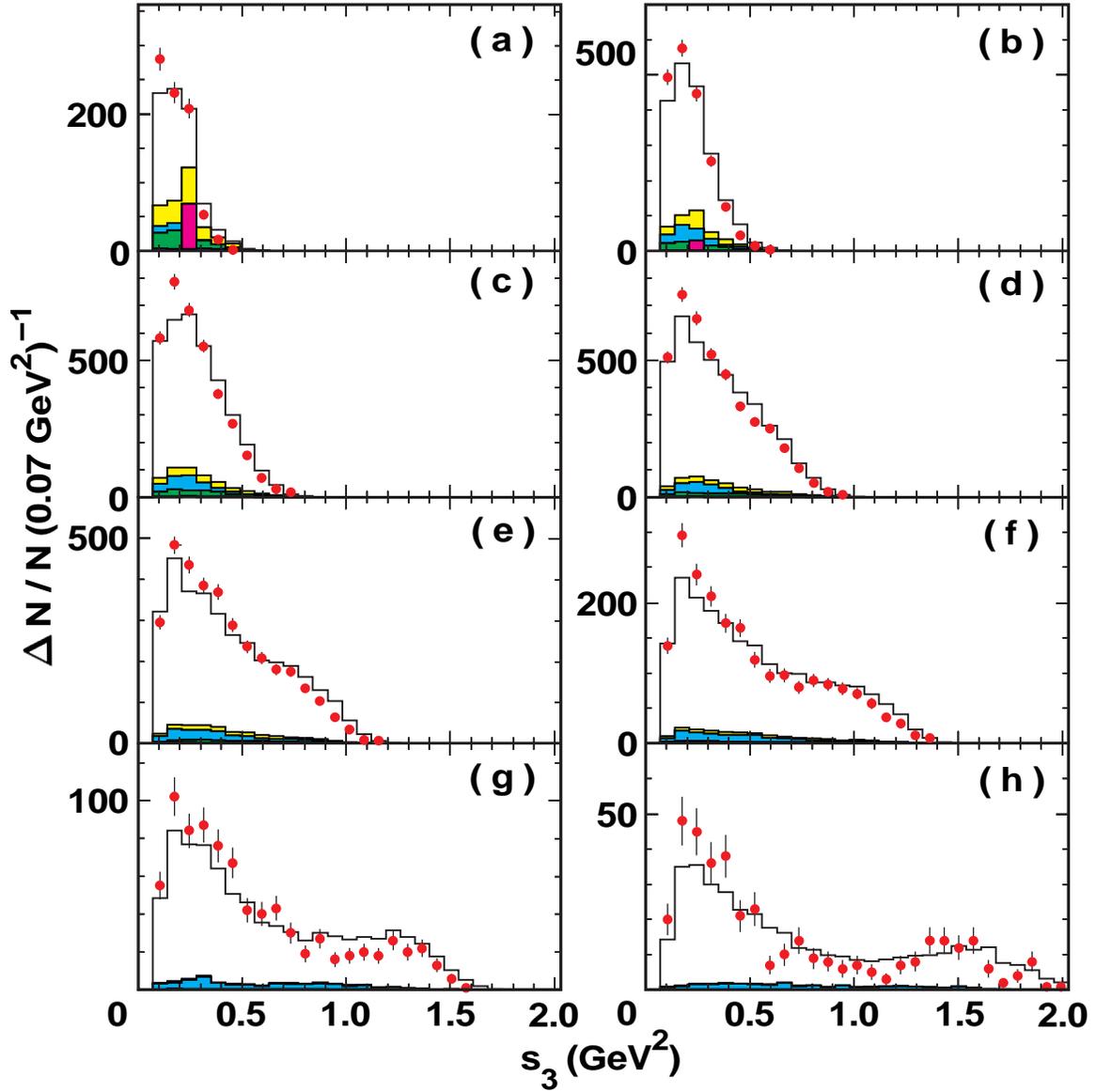}
\caption[]{Dalitz plot projections:  
distributions in squared $\pi^0\pi^0$ mass $s_3$ (one entry per event),
plotted for slices of $\sqrt{s}$ as in Fig.~\ref{s3_fitb}.  
The data are represented by the filled points. The solid
line is the result from a modified version of the nominal fit 
in which amplitudes involving isoscalar mesons 
($\sigma\pi$, $f_0\pi$ and $f_2\pi$) have been omitted.  
}
\label{fig:s3_noiso}
\end{figure}

To assess the statistical significances of individual amplitudes in the 
nominal substructure fit summarized in Table~\ref{tab:result_r00}, 
we successively repeated the fit, each time with one amplitude omitted.  
In view of the large contribution from amplitudes involving 
isoscalar mesons, we have also performed a fit in which all three of 
these have been omitted.  The Dalitz plot projections from this fit 
are plotted in slices of $\sqrt{s}$ in Figs.~\ref{fig:s1s2_noiso} and 
\ref{fig:s3_noiso}, overlaid on the data distributions.  The 
agreement with the data is visibly worse than that seen in the nominal 
fit with all amplitudes (see Figs.~\ref{s1s2_fitb} and \ref{s3_fitb}).

\section{RESULTS FROM FITS WITH FINITE MESON RADII}
\label{a-mesonradius}

  Treatment of the $a_1$ as being point-like results in unphysical behavior 
of the running width $\Gamma^{a_1}_{tot}(s)$ at large values of $s$.  
The effect of the form factor $F_{R_i}(k_i)$ defined in 
Eqn.~\ref{eq:mesonradius} 
is to damp out this behavior, although the Gaussian form itself is somewhat 
{\sl ad hoc}.  This form factor affects both the substructure analysis 
and the $3\pi$ mass spectrum analysis.  Although the data at present do not 
show much sensitivity to the value of the meson size $R$, we note that 
both analyses prefer values of $R\sim 1.4\,\rm{GeV}^{-1}$.  In this 
appendix we report the results from fits using non-zero values for $R$.

\subsection{Substructure fits}

In Table~\ref{tab:subfitr14}, we give the results from the substructure 
fit with $R_i = 1.4\,\rm{GeV}^{-1}$.  These results are in qualitative 
agreement with those from the nominal fit.  
\begin{table}[hbpt]
\begin{center}
\caption[]{\small Results from substructure fit with meson radii 
$R_i = 1.4\mbox{ GeV}^{-1}$.} 
\label{tab:subfitr14}
\begin{tabular}{lccccc}
%
&   & Signif. 
& $\vert \beta_i \vert$ 
& $\phi_{\beta_i} /\pi$ 
& ${\cal B}$ fraction ($\%$) 
%
\\ \hline
%
$\rho$         & $s$-wave & 
--  & $ 1.00 $ & $ 0.0  $ & $ 69.11 $ \\
$ \rho(1450) $ & $s$-wave &  
$0.6\sigma   $ & $0.03\pm 0.06$&$\phantom{-}0.92\pm 0.58 $ & $0.02\pm 0.08$ \\
$ \rho       $ & $d$-wave &
$5.4\sigma   $ & $0.36\pm 0.09$&$-0.12\pm 0.11 $ & $0.28\pm 0.13$ \\
$ \rho(1450) $ & $d$-wave &
$3.7\sigma   $ & $0.78\pm 0.28$&$\phantom{-}0.41\pm 0.14 $ & $0.28\pm 0.20$ \\
$ f_2 (1270) $ & $p$-wave &
$4.6\sigma   $ & $0.66\pm 0.14$&$\phantom{-}0.64\pm 0.10$ & $0.11\pm 0.05$ \\
$ \sigma     $ & $p$-wave &
$7.9\sigma   $ & $2.13\pm 0.27$&$\phantom{-}0.27\pm 0.03$ & $14.31\pm 3.66$\\
$ f_0 (1370) $ & $p$-wave &
$6.1\sigma   $ & $0.77\pm 0.13$&$-0.47\pm 0.05$ & $3.69\pm 1.21$ \\
%
%
\end{tabular}
\end{center}
\end{table}

\subsection{Fits to \boldmath $M_{3\pi}$ }

Fits to the $3\pi$ mass spectrum were performed with non-zero values 
for $R_i$ with the assumptions of constant and running $a_1$ masses.
These studies differed from the nominal fit 
in Fig.~\ref{fig:all_m3pi_n} in two additional ways.  First, the fitting 
range included two additional bins in the high mass region, extending 
up to 1.775 MeV.  Second, so as to account for possible systematic 
effects, the acceptance and background corrections were allowed 
to vary within reasonable amounts by adding terms to the $\chi^2$ constraining 
their deviations from the nominal assuming these to be Gaussian distributed 
(see Section~\ref{ss-masssys}).

Results for fits assuming constant and running $a_1$ masses are presented 
in Tables~\ref{tab:mass_rvar_c} and~\ref{tab:mass_rvar_r} respectively.  
Similar trends are observed in fits to just the lepton-tagged sample, as 
well as in fits that include an $a_1^\prime$ contribution.  The 
final column in the tables, labeled $\sqrt{\chi^2_c}$, denotes the 
square root of the contribution of the acceptance and background correction 
constraints to the total $\chi^2$ of the fit.  In both constant and running 
mass scenarios, the $a_1$ pole width is strongly affected by the presence 
of the form factor accounting for the size of the $a_1$ meson, and by the 
value of $R$.  

\begin{figure}
  \centering
  \leavevmode
  \epsfysize=16.0cm
  \epsfbox{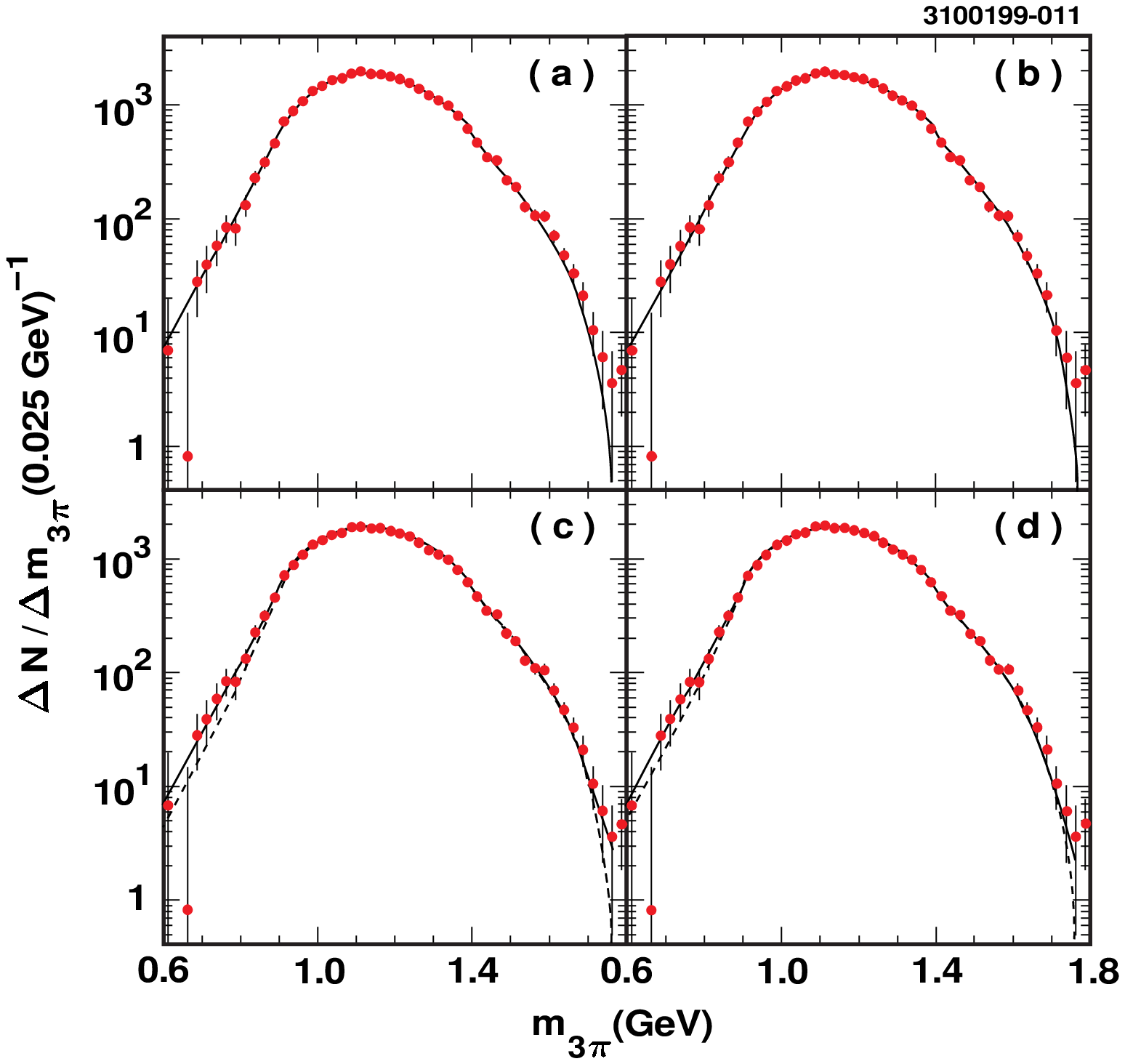}
\caption[]{Background-subtracted efficiency-corrected $3\pi$ mass
spectrum (points with error bars) with fit functions overlaid. 
The solid curves in plots (a) and (b) 
represent the fit function with $R =1.2\mbox{ GeV}^{-1}$, assuming a constant 
mass, excluding (a) and including (b) the effect of the $f_0(980)\pi$ threshold 
on $\Gamma^{a_1}_{tot}(s)$.  The solid curves in plots (c) and (d) represent 
the fit function with $R =1.4\mbox{ GeV}^{-1}$, assuming a running mass.  
The effect of the $f_0(980)\pi$ threshold on $\Gamma^{a_1}_{tot}(s)$ and 
$m_{a_1}(s)$ is excluded in (c), and included in (d).  
The dotted curves represent the corresponding fit functions 
without the deviations in the background subtraction and efficiency
correction returned by the fits.}
\label{fig:mass_rvar}
\end{figure}

The preferred fits, with $R=1.2\;\mbox{GeV}^{-1}$ for the constant mass case 
and $R=1.4\;\mbox{GeV}^{-1}$ for the running mass case are plotted in 
Figures~\ref{fig:mass_rvar}(a) and~(c), respectively.  We have also performed 
these fits including turn-on of the $f_0(980)\pi$ channel, the results of 
which are plotted in Figures~\ref{fig:mass_rvar}(b) and~(d) for the two cases.
\begin{table}
\begin{center}
\caption[]{Results of $3\pi$ mass spectrum fits 
           with different input values of $a_1$ size parameter $R$, 
           assuming constant $a_1$ mass.}
\label{tab:mass_rvar_c}
\begin{tabular}{cccccc}
$R$                  & $\chi^2 / n_{dof}$ & 
  $m_{0a_1}$ & $\Gamma_{0a_1}$ & ${\cal B} (K^\star K ) $  & $\sqrt{\chi^2_c}$ \\
$[\mbox{GeV}^{-1}]$  &                     & 
$[\mbox{GeV}]$ & $[\mbox{GeV}] $  & $[\%]$ &  \\ \hline
$0.0$   & 40.9/43 & $1.330\pm 0.011$ & $0.811\pm 0.042$ & $3.3\pm 0.6$ & $0.32$  \\
$1.0$   & 38.9/43 & $1.289\pm 0.008$ & $0.653\pm 0.025$ & $3.4\pm 0.6$ & $0.32$  \\
$1.2$   & 38.6/43 & $1.285\pm 0.007$ & $0.619\pm 0.021$ & $3.5\pm 0.6$ & $0.31$  \\
$1.3$   & 39.3/43 & $1.279\pm 0.006$ & $0.597\pm 0.019$ & $3.4\pm 0.7$ & $0.33$  \\
$1.4$   & 39.9/43 & $1.274\pm 0.006$ & $0.578\pm 0.017$ & $3.2\pm 0.7$ & $0.37$  \\
$1.5$   & 42.9/43 & $1.269\pm 0.005$ & $0.558\pm 0.015$ & $3.0\pm 0.7$ & $0.38$  \\
$1.6$   & 45.5/43 & $1.263\pm 0.005$ & $0.538\pm 0.014$ & $2.8\pm 0.7$ & $0.44$  \\
\end{tabular}
\end{center}
\end{table}
\begin{table}
\begin{center}
\caption[]{Results of $3\pi$ mass spectrum fits 
           with different input values of $a_1$ size parameter $R$, 
           assuming running $a_1$ mass.}
\label{tab:mass_rvar_r}
\begin{tabular}{cccccc}
$R$                  & $\chi^2 / n_{dof}$ & 
  $m_{0a_1}$ & $\Gamma_{0a_1}$ & ${\cal B} (K^\star K ) $  & $\sqrt{\chi^2_c}$ \\
$[\mbox{GeV}^{-1}]$  &                     & 
$[\mbox{GeV}]$ & $[\mbox{GeV}] $  & $[\%]$ &  \\ \hline
$1.2$ & 39.6/43 & $1.298\pm 0.007$ & $1.200\pm 0.100$ & $3.3\pm 0.7$ & $0.77$  \\
$1.3$ & 39.3/43 & $1.294\pm 0.006$ & $0.822\pm 0.047$ & $3.3\pm 0.7$ & $0.78$  \\
$1.4$ & 36.7/43 & $1.288\pm 0.006$ & $0.667\pm 0.031$ & $3.3\pm 0.7$ & $0.71$  \\
$1.5$ & 41.9/43 & $1.284\pm 0.006$ & $0.558\pm 0.021$ & $3.2\pm 0.7$ & $0.79$  \\
$1.6$ & 54.4/43 & $1.279\pm 0.005$ & $0.493\pm 0.016$ & $3.0\pm 0.7$ & $0.77$  \\
\end{tabular}
\end{center}
\end{table}

\begin{figure}
  \centering
  \leavevmode
  \epsfysize=8.0cm
  \epsfbox{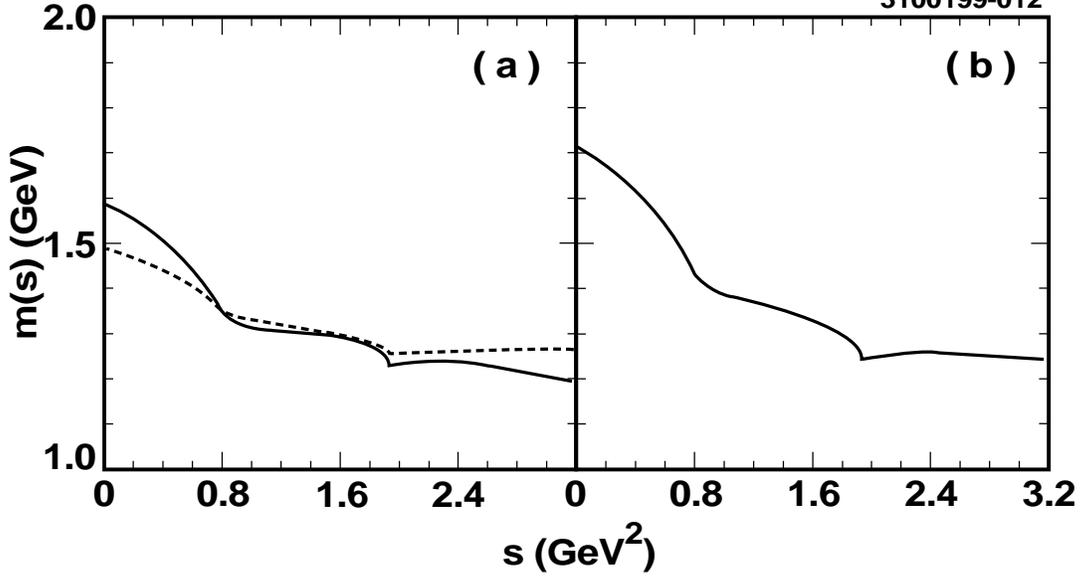}
  \caption[]{The solid curve in plot (a) shows the running mass used in the fit
  shown in Fig.~\ref{fig:mass_rvar}(c).  The dotted curve illustrates the 
  effect of adding the $f_0 (980)\pi$ threshold, with $\Gamma_{3\pi}(s)$ and 
  $\Gamma_{K^\star K}$ unchanged.   
  Plot (b) shows the running mass
  used in the fit shown in Fig.~\ref{fig:mass_rvar}(d), where the relative $K^*K$ 
  and $f_0(980)\pi$ amplitudes are determined from the fit.  }
  \label{fig:all_rmass}
\end{figure}

The parametrizations of the $\sqrt{s}$-dependence of the $a_1$ mass entering 
the fits shown in Figs.~\ref{fig:mass_rvar}(c) and~(d) are plotted in 
Fig.~\ref{fig:all_rmass}.  The overall mass shift function depends on 
the relative amplitudes for the $K^*K$ and (in Fig.~\ref{fig:mass_rvar}(d)) 
$f_0(980)\pi$ channels which are fit parameters, however the shapes of 
the contributions to $m_{a_1}(s)$ from these channels are 
determined as described in Section~\ref{ss-modmass}.  The effect of 
successively including thresholds is illustrated by the dotted curve in 
Fig.~\ref{fig:all_rmass}(a), in which an {\sl ad hoc} $f_0(980)\pi$ contribution 
is added, assuming unchanged couplings to the other channels.  As noted 
earlier, the effect is to flatten the dependence of $m_{a_1}(s)$. 

Satisfactory fits are obtained with both constant and running $a_1$ masses. 
Functions assuming a running mass yield fits with smaller $\chi^2$ values than 
those with a constant mass.  However, the running mass fits also prefer a larger 
distortion of the background and acceptance corrections, as indicated by 
the larger values of $\sqrt{\chi^2_c}$ in Table~\ref{tab:mass_rvar_r} and 
by the deviations of the dotted curves in Figs.~\ref{fig:mass_rvar}(c) 
and~(d) from the corresponding solid curves.  


\clearpage
%
%
%


\end{document}